\documentclass[aps,PRApplied,reprint,superscriptaddress]{revtex4-2}

\usepackage{graphicx}
\usepackage{color}
\usepackage[english]{babel}
\usepackage{rotating}
\usepackage{braket}
\usepackage{multirow}
\usepackage{makecell}
\usepackage{longtable}

\graphicspath{{./img/}}

\begin{document}

\title{Machine Learning-Assisted Manipulation and Readout of Molecular Spin Qubits}

\author{Claudio Bonizzoni}
\email[mail to:]{ claudio.bonizzoni@unimore.it}
\affiliation{Dipartimento di Scienze Fisiche, Informatiche e Matematiche Universit\`a di Modena e Reggio Emilia, via G. Campi 213/A, 41125, Modena, Italy}
\affiliation{CNR Istituto Nanoscienze, Centro S3, via G. Campi 213/A, 41125, Modena, Italy.}
\author{Mirco Tincani}
\affiliation{Dipartimento di Scienze Fisiche, Informatiche e Matematiche Universit\`a di Modena e Reggio Emilia, via G. Campi 213/A, 41125, Modena, Italy}
\author{Fabio Santanni}
\affiliation{Dipartimento di Chimica Ugo Schiff, via della Lastruccia 3, 50019, Sesto Fiorentino (FI), Italy}
\author{Marco Affronte}
\affiliation{Dipartimento di Scienze Fisiche, Informatiche e Matematiche Universit\`a di Modena e Reggio Emilia, via G. Campi 213/A, 41125, Modena, Italy}
\affiliation{CNR Istituto Nanoscienze, Centro S3, via G. Campi 213/A, 41125, Modena, Italy.}

\date{\today}

\begin{abstract}
Machine Learning finds application in the quantum control and readout of qubits. In this work we apply Artificial Neural Networks to assist the manipulation and the readout of a prototypical molecular spin qubit - an Oxovanadium(IV) moiety - in two experiments designed to test the amplitude and the phase recognition, respectively. We first successfully use an artificial network to analyze the output of a Storage/Retrieval protocol with four input pulses to recognize the echo positions and, with further post selection on the results, to infer the initial input pulse sequence. We then apply an Artificial Neural Network to ascertain the phase of the experimentally measured Hahn echo, showing that it is possible to correctly detect its phase and to recognize additional single-pulse phase shifts added during manipulation.
\end{abstract}


\maketitle

\section{Introduction}
\label{sec.intro}

Machine Learning (ML) methods are extremely versatile and flexible algorithms finding large applications in a continuously  increasing number of fields \cite{carleoREVMODPHYS2019,schmidtNPJCOMPMAT2019,cuiACSSENSORS2020,striethkalthoffCHEMSOCREV2020,butlerNATURE2018,materJCHEMINFMOD2019}. Merging ML with the latest, state of the art, progresses of quantum technologies has contributed to develop algorithms working on quantum states or in which genuine quantum features are used to enhance the capability of the algorithms themselves \cite{biamonteNATURE2017,khanIEEEACCESS2020,tensorflowquantum}. ML has found application in quantum optics and metrology \cite{polinoAVSQUASCI2020} \textit{e.g.} in phase estimation problems \cite{rambhatlaPHYSREVRES2020}, in sensor calibration \cite{ciminiPRAPPL2021} and in the readout of trapped-ions qubits \cite{seifJOPB2018}. Semicondutor-based quantum dots have benefited from ML in their fabrication processes \cite{meiAPL2021}, in the automatic search and tuning of their working points \cite{kalantreNPJQUANTINF2019,moonNATCOMM2020,teskeAPL2019} and in their measurement \cite{lennonNPJQUANTINF2019}. Similar advantages have been recently reported also on the design \cite{Menke2021}, the quantum optimal control \cite{sivakPRX2022} and the readout \cite{koolstraPRX2022,liuMACHLEARNSCITECH2020,lienhardPHDTHESIS2021} of superconducting qubits. ML techniques have been recently successfully implemented on electronic and nuclear spins resonance experiments both at microwave (MW) and radio (RF) frequency. Along this line, promising results were obtained on Nitrogen-Vacancy (NV) centers in diamonds to enhance and optimize the contrast of their optical readout \cite{qianAPL2021,SantagatiPRX2019,fujisakuACSMEASUAU2021}, to sense their surrounding nuclear spin bath \cite{jungNPJQUANTINF2021} or for 2D imaging \cite{kongNPJQINF2020}. Nuclear Magnetic Resonance-based quantum processors \cite{WangADVQUANTTECH2022,Manu2012}, the recognition of electronic spin echoes from noise background \cite{polulyakhTECHPHYSLETT2019}, and the design and optimization of pulse sequences in Nuclear Magnetic Resonance spectroscopy \cite{Manu2015,Bechmann2013} have been also investigated. ML has been applied also to HYSCORE sequences to analyze correlation of hyperfine signal \cite{taguchiJOURNPHYSCHEM2019}. To the best of our knowledge, most of ML approaches reported are based on Artificial Neural Networks, that are networks of fundamental units (nodes or neurons) in which learning and execution of a task aim to mimic the behaviour of the human brain \cite{haganebook}.

Molecular spin qubits have been recently shown to have long coherence times over a wide range of temperature \cite{zadroznyACSCentrSci2015,atzoritesiJACS2016,atzorimorraJACS2016,baderNATCOMM2014,baderChemComm2016,shiddiqNAT2016,gaitaarinoNATCHEM2019,gimenezCHEMSCI2020}. The viability of their integration into hybrid quantum circuits at MW frequency and in solid-state quantum technologies has been also demonstrated both in the Continuous Wave (CW) \cite{bonizzoniAdvPhys2018,bonizzoniSCIREP2017,ghirriPRA2016,ghirriAPL2015,mergenthalerPRL2017,gimenoACSNANO2020,gimenoCHEMSCI2021} as well as in the Pulsed Wave (PW) regime \cite{bonizzoniNPJQUANT2020,lenzADVMAT2021,jenkinsPRB2017} of excitation. For instance, molecular spin qubits were found to reach the coherent spin-photon coupling \cite{bonizzoniAdvPhys2018,bonizzoniSCIREP2017,ghirriPRA2016,ghirriAPL2015,mergenthalerPRL2017,lenzCHEMCOMM2020} and to be suitable as quantum memories for information \cite{bonizzoniNPJQUANT2020,lenzADVMAT2021}. Their readout in the dispersive, non-resonant regime has been also reported \cite{bonizzoniADVQTECH2021}. The implementation of quantum algorithms on single molecules in a spin transistor geometry has been also experimentally demonstrated \cite{thieleSCIENCE2014,godfrinNPJQUANTINF2018,godfrinPRL2017}. Several protocols (\textit{i.e.} PW sequences at MW and RF frequency) were proposed to encode and process information using quantum states built from molecular spin qudits \cite{castroPRAPPL2022,gaitaarinoNATCHEM2019,aguilaJACS2014,ferrandosoriaNatComm2016}, with also the potential advantages of an already-embedded quantum error correction for gate operations \cite{chizziniPHYSCHEMCHEMPHYS2022,Chiesa2021,Chiesa2020,macalusoCHEMSCI2020} and of a multiqubit dispersive readout \cite{gomezleonPRAPPL2022}. Along these lines, however, no ML algorithm has been developed for the manipulation and readout of molecular spin qubits yet.

In this work we test Artificial Neural Networks (hereafter, ANNs) to assist the resonant readout and manipulation of an oxovanadium(IV) complex, VO(TPP) (where TPP$^{2-}$ is the Tetraphenyl porphyrinate ligand), a molecular spin qubit on which we have recently demonstrated the coherent manipulation and the implementation of a Storage/Retrieval protocol when embedded into a planar superconducting MW resonator \cite{bonizzoniNPJQUANT2020}. Thanks to its electronic spin $S=1/2$,  this system constitutes a prototypical spin qubit for testing ML algorithms. We first revisit the Storage/Retrieval protocol \cite{bonizzoniNPJQUANT2020} to codify into the ensemble all the $2^{4}=16$ sequences obtainable from 4 binary digits (decimal numbers from 0 up to 15), and we show that an ANN can be used to recognize each of the resulting weak output echo/es signal/s without any prior knowledge on their number and position. A further post selection performed with an unsupervised ML method (K-Means clustering) allows us to successfully infer the initial input bit sequence with high ($\geq\,97\,$\%) accuracy.
We then test the readout of the phase of an optimal Hahn echo using an ANN. We show that it is possible to correctly determine the initial phase of the spin precession from the analysis of raw measured quadrature outputs. The ANN is found to recognize the effects of additional single-pulse phase shifts introduced in the spin precession during its initialization ($\pi/2$ pulse) or during its refocusing ($\pi$ pulse). These results demonstrate the possibility to use ANNs to assist the readout of the amplitude or phase of (molecular) spin qubits, a key aspect for the implementation of gate operations. Our approach can find more general application to all systems showing quantum coherence in the form of refocusing echo signal/s and it can also be further extended down to the case when quantum regime of the driving electromagnetic radiation is achieved. Potential systems along this line can be superconducting qubits \cite{Kjaergaard2020,krantzAPL2019} and solid-state spin qubits based on semiconductor quantum dots \cite{chatterjeeNATPHYSREV2021} or on diluted magnetic centers \cite{mortonJMR2018}.

\section{Experimental Methods}

\subsection{Experimental Set Up and PW Sequences}
\label{sec_setup}

We use a 2\% doped polycrystalline powder of VO(TPP) in its isostructural diamagnetic analog TiO(TPP). Each molecule has an electronic spin $S\,=\,1/2$ ground state and an additional hyperfine splitting given by the interaction with the $I=7/2$ nuclear spin of the $^{51}$V ion (natural abundance: 99.75\%). The magnetic properties and the electron spin resonance spectroscopy of this molecule have been previously reported in \cite{yamabayashiJACS2018}. We perform our experiments by placing the sample on a superconducting coplanar resonator ($\nu_{0}\approx\,6.91$ GHz) made out of superconducting YBa$_2$Cu$_3$O$_7$ (YBCO) films on a Sapphire substrate, as described in \cite{bonizzoniNPJQUANT2020,bonizzoniSCIREP2017,ghirriAPL2015}. The CW and PW microwave spectroscopy of VO(TPP) through the resonator has been previously reported in \cite{bonizzoniNPJQUANT2020}. The sample and the resonator are cooled-down to 4 K into a commercial Quantum Design Physical Properties Measurement System (QD PPMS), which is also used to apply the external static magnetic field \cite{bonizzoniNPJQUANT2020,bonizzoniAdvPhys2018}.\\ 

Our Storage/Retrieval protocol consists of a train of 4 weak MW pulses equally-spaced in time, with duration $t_{p}=40\,$ ns and interpulse delay $t_{d}=300\,$ ns. A $\pi$ pulse with duration $t_{\pi}=190\,$ ns is sent after a delay $\tau=1200\,$ ns with respect to the last pulse of the input train. A relaxation time $t_{relax}=15\,$ms is added at the end of the sequence to avoid sample saturation. The refocusing occurs after an additional delay $\tau$ with respect to the $\pi$ pulse, giving a train of weak output echoes. This protocol allows us to use the spin ensemble as a temporary memory for information \cite{grezesPhysRevX2014,ranjanPRL2020,bonizzoniNPJQUANT2020}. In this work we exploit the 4 input pulses to codify into the ensemble the binary sequences corresponding to 16 decimal numbers (from 0 to 15, \textit{i.e.} from 0000 to 1111). An input pulse ON corresponds to the classical logical bit 1 (visible output echo), while a pulse OFF gives the classical logical bit 0 (no output echo). We denote the position of each pulse in the input train with the index $i=1,\dots,4$, according to their order of generation. In other words, the index $i$ will give the Storage order into the ensemble and the weight of the bit from the most significant to the least significant one.\\  

The Hahn echo sequence used in the phase recognition experiments consists of two pulses spaced by an interpulse delay $\tau=750\,$ ns. Their durations are $t_{\pi/2}=145\,$ ns and $t_{\pi}=180\,$ ns, respectively. A relaxation time $t_{relax}=20\,$ms is added at the end of the sequence to avoid sample saturation. The phase of each input pulse can be controlled independently by the AWG. The phase generation and the whole phase-sweep acquisition is controlled by a home-written Python script. We test a similar phase recognition on additional data sets measured in slightly different experimental conditions, in which $t_{\pi/2}=120\,$ ns, $t_{\pi}=290\,$ and $\tau=800\,$ ns (see Supplementary Information).

\subsection{Machine Learning Methods}
\label{sec_ml}

\begin{figure}[t]
\centering
\includegraphics[width=0.5\textwidth]{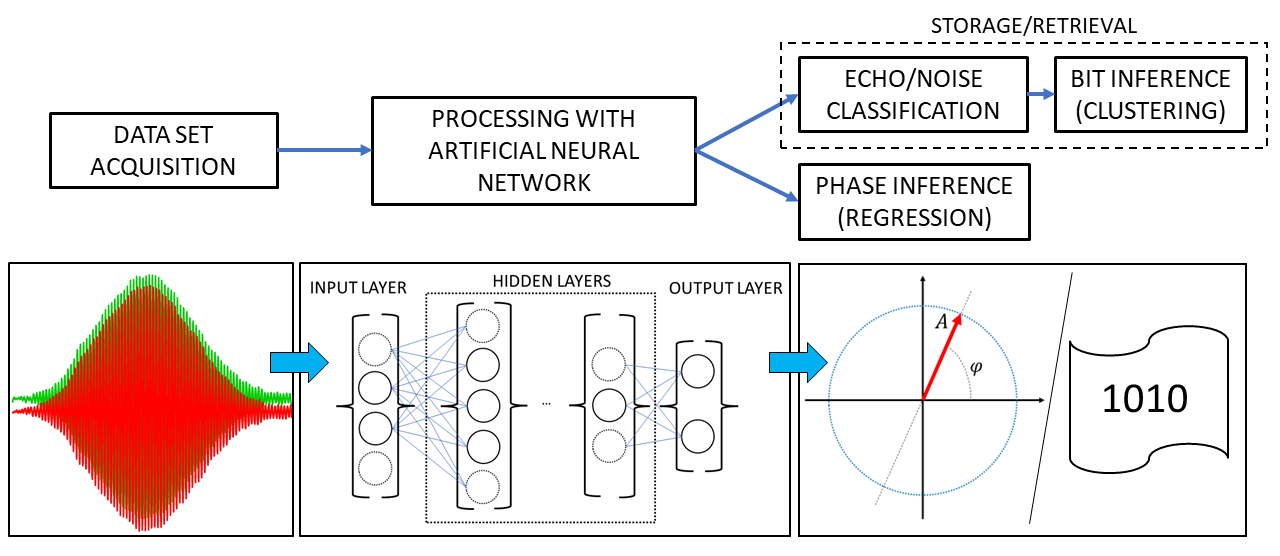}
\caption{Summary of the experimental workflow. The flow chart on top summarizes the main steps of our method, from data acquisition, to signal processing and to the execution of the final task (classification or regression). The three sketches at bottom help in recognizing different steps (raw data acquisition, processing, prediction). We use off-line methods, in which acquisition and processing with ANN are done in two distinct steps. Details on training procedures are given in Supplementary Information.}
\label{fig_ml_ann}
\end{figure}

We use Artificial Neural Networks (ANNs) \cite{haganebook} for the experiments described in this work. Each Network has been implemented using the Keras and Tensorflow Python packages \cite{tensorflow,keras} within home-written Python scripts. We used sequential networks characterized by dense layers. The topology and the parameters of each network have been optimized with the Python Talos package \cite{talos} before the final experiments. In this work we test off-line methods, in which experimental acquisition of the data and ML processing are done in two different sequential steps. The experimental workflow for the analysis of raw data is briefly summarized in Fig. \ref{fig_ml_ann}, while the training procedure is described in Supplementary Information.   

\subsection{Echo Recognition in Storage/Retrieval}
We design the ANN for a classification problem, in which two different labels (1 = echo, 0 = noise) are assigned to an input signal. The input and the output traces are normalized before training and before using the ANN for predicitions. The test loss found after optimization is $J=3\cdot10^{-2}$ with an accuracy of $99.2\,\%$. The output of the network is a 2D vector of normalized values, $(p_{e},p_{n})$, which results from the propagation of the input across the network. These values represent the confidence of the input to be an echo ($p_{e}$) or noise ($p_{n}$), respectively, and can be though as equivalent to two probability values. In the following, since $p_{n}=1-p_{e}$, we restrict our analysis to $p_{e}$ and we will refer to it as \emph{echo probability}. Further details about the ANN and the training data set used are given in Supplementary Information. The training time required by the optimized ANN described above and with our data sets was found to be 90 s, while the one required for predicting output values from an unknown data set was found to be always below 1 ms, with typical values of $\approx 500 \mu$s. 

The bit inference of Sec. \ref{sec_ml_amplitude} is done by performing an additional post-selection on the values of $p_{e}$ obtained for each trace. This is done with the K-Means clustering method of the Scikit-Learn Python package\cite{scikitlearn}. Briefly, each trace is divided into four equally-spaced windows. The K-means clustering is then applied to each window to assign the data points to two clusters (accounting for two possible logical outcomes with unknown label). The final probability for each window is assigned using the value associated to the centroid of the cluster with the larger number of points. This post-selection gives the probability values used to calculate the Fidelity in Fig. \ref{fig_stor_retr_fid} through Eq. \ref{eq_fidelity}. Examples of clusterization obtained from the results of Fig. \ref{fig_stor_retr_echo} are shown in Supplementary Information.

\subsection{Phase Recognition from Hahn's Echo}
We design the ANN for a regression problem, in which two raw traces given by the I and Q output ports of the detection mixer (which are not the echo quadratures but their combinations, see Supplementary Information) are used to predict their corresponding phase value. Here, only a small window of the traces corresponding to few periods (40+40 points using a symmetric time window taken around the maxima of the echo in the I channel) of the down converted carrier frequency (90 MHz) is given as input instead of the whole echo trace. This allows us to reduce the size of the ANN and has been checked to not affect the results. The test loss resulting after network optimization is $J=4\cdot10^{-6}$. The training time required by the optimized ANN was 49 s, while the typical one required for making predictions was $\approx 500 \mu$s. Further details on the ANN and on the training data set are reported in Supplementary Information. In this work, we focus on the average phase value of the Hahn echo signal but we remark that, with the proper modification, our approach can be extended also to the instantaneous phase value (see Supplementary Information for details).

\section{Results}
\label{sec_results}

\subsection{Machine Learning-Assisted Echo Recognition}
\label{sec_ml_amplitude}

\begin{figure}[h!]
\centering
\includegraphics[width=0.24\textwidth]{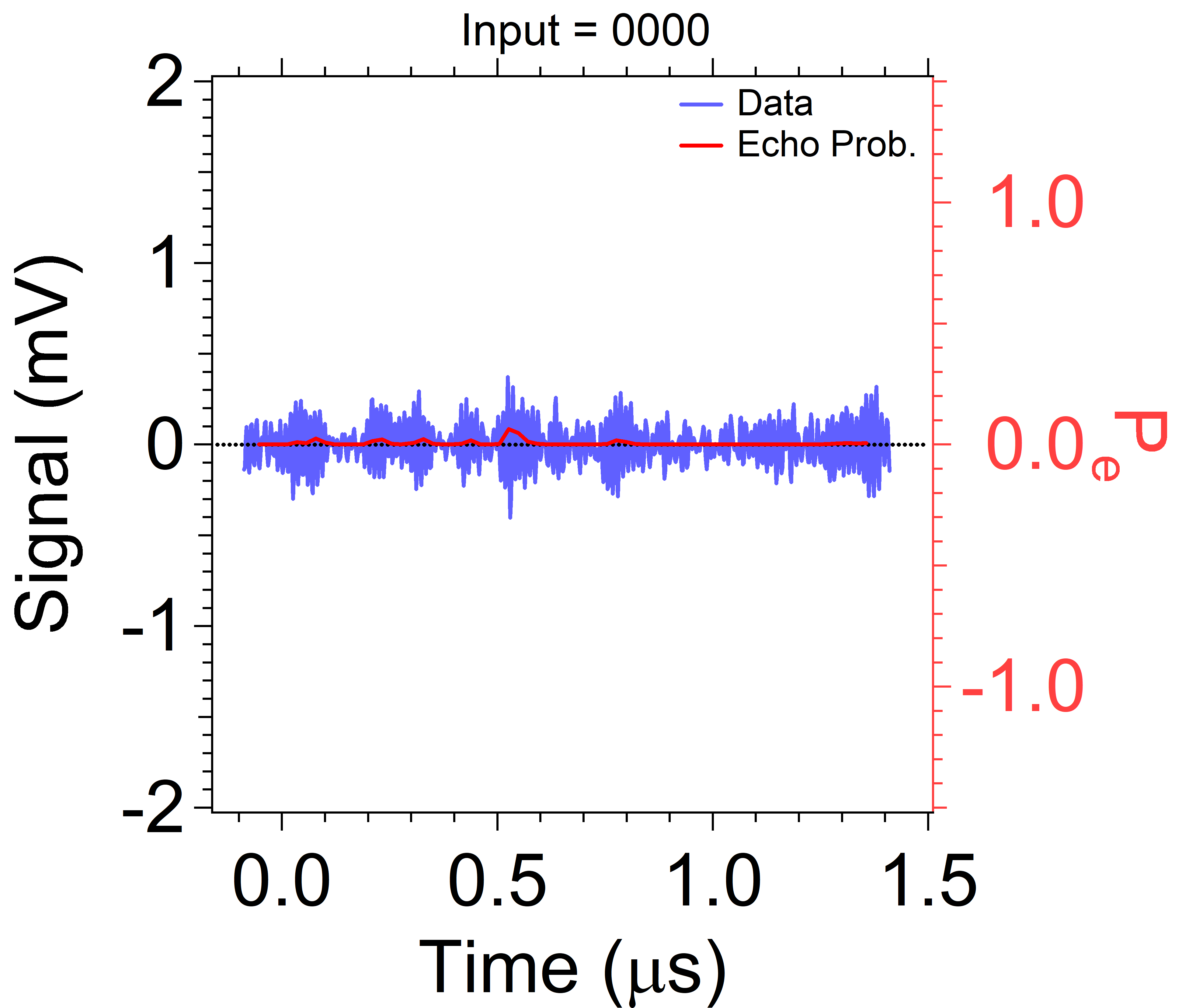}
\includegraphics[width=0.24\textwidth]{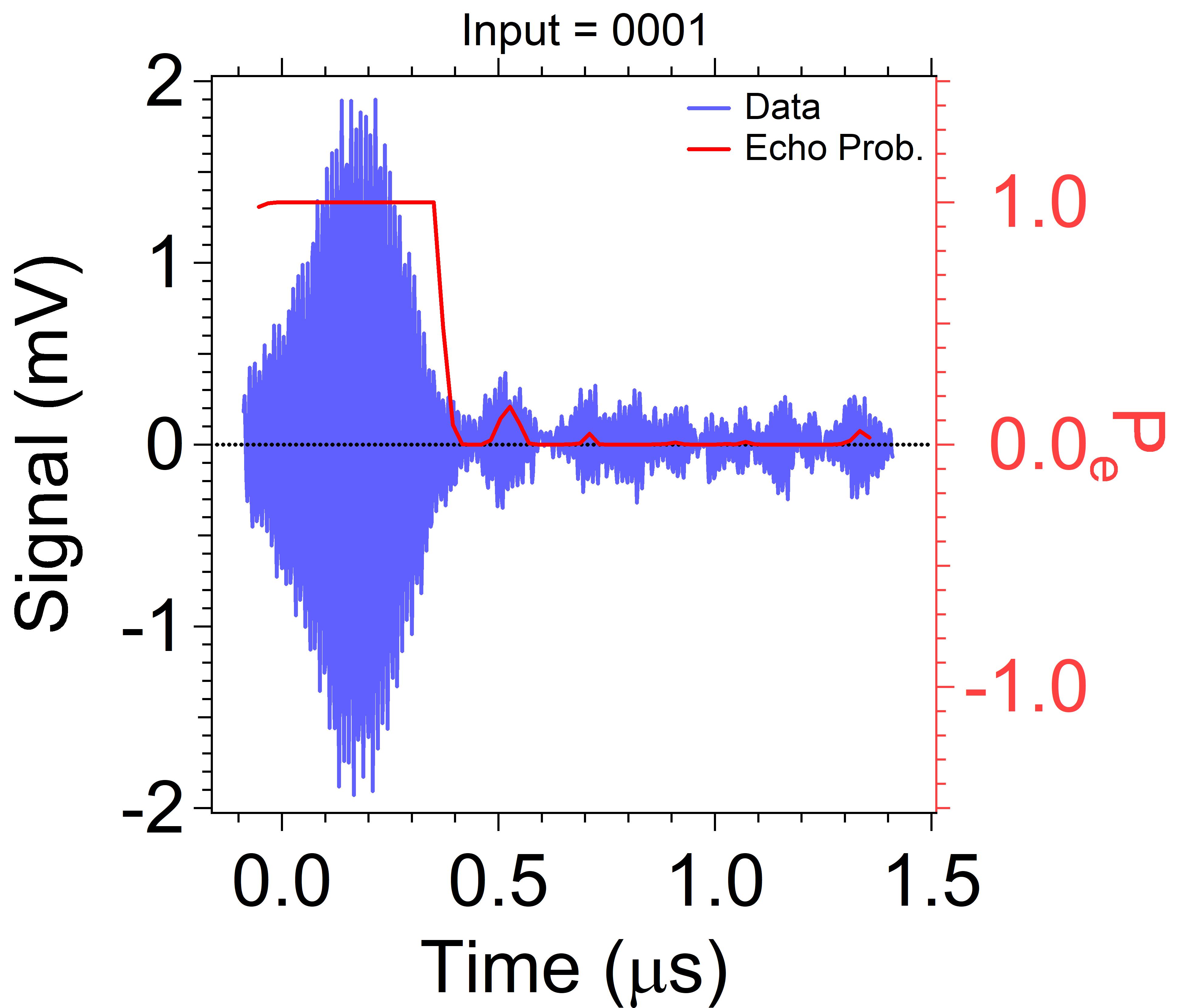}
\includegraphics[width=0.24\textwidth]{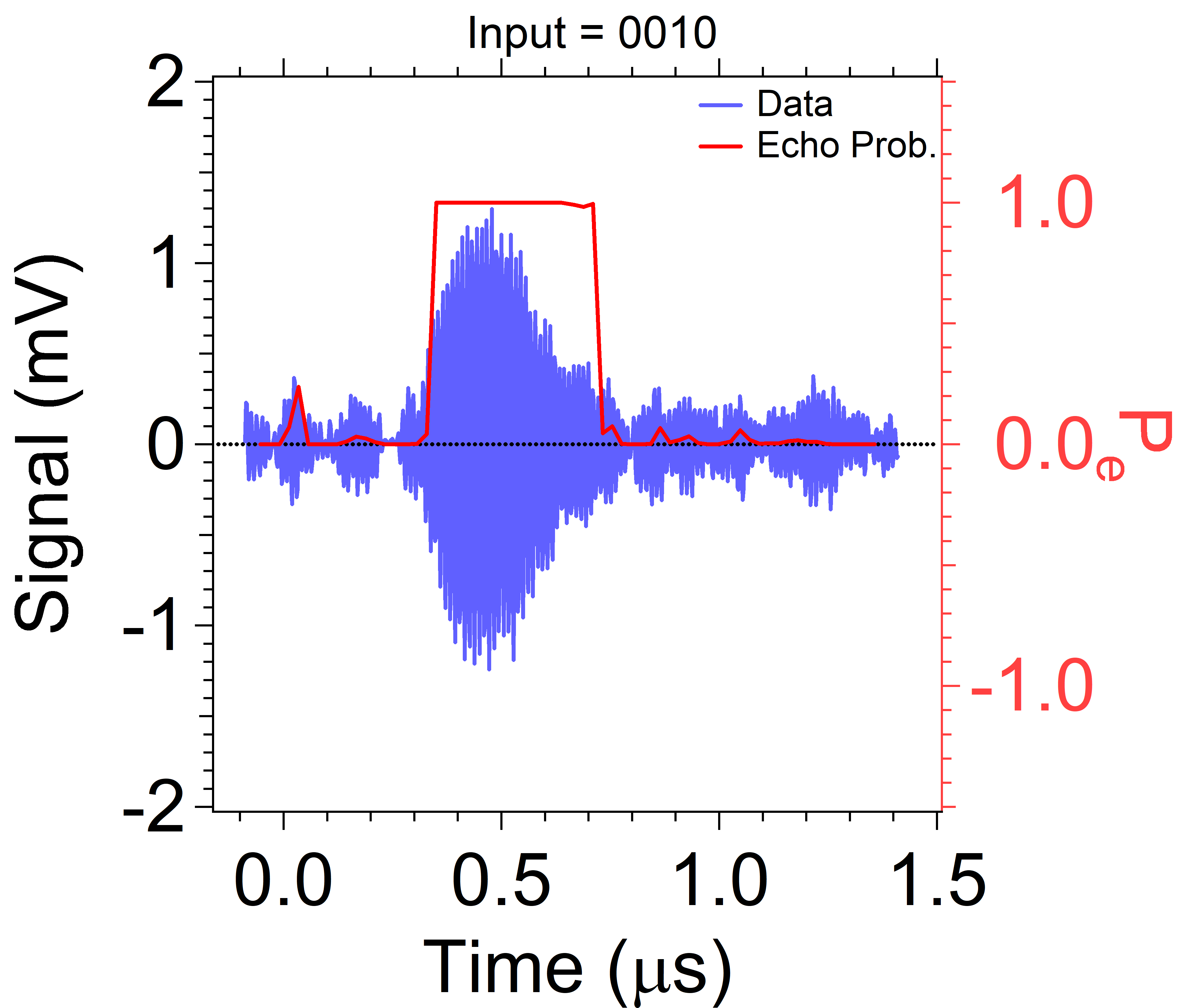}
\includegraphics[width=0.24\textwidth]{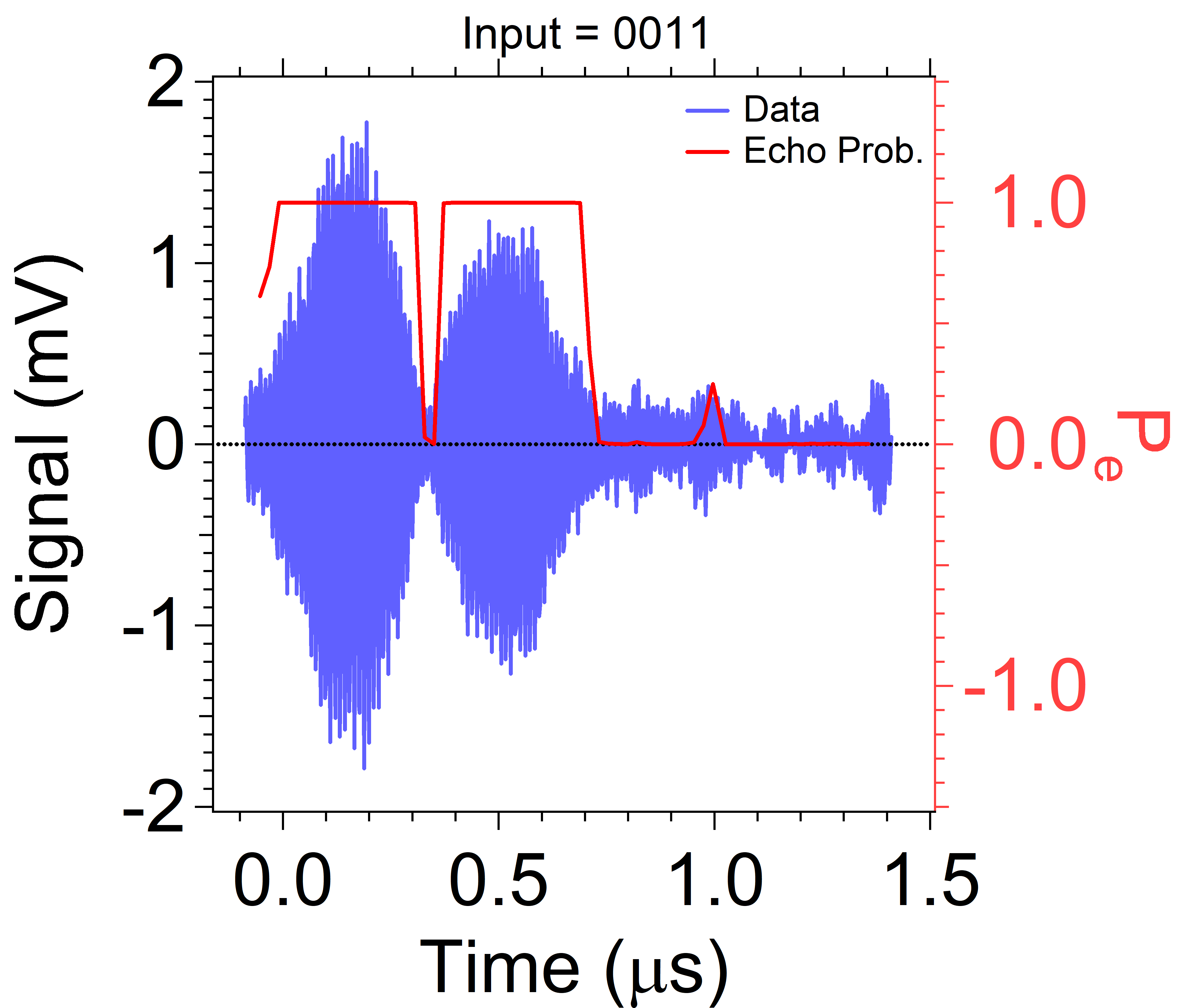}\\
\includegraphics[width=0.24\textwidth]{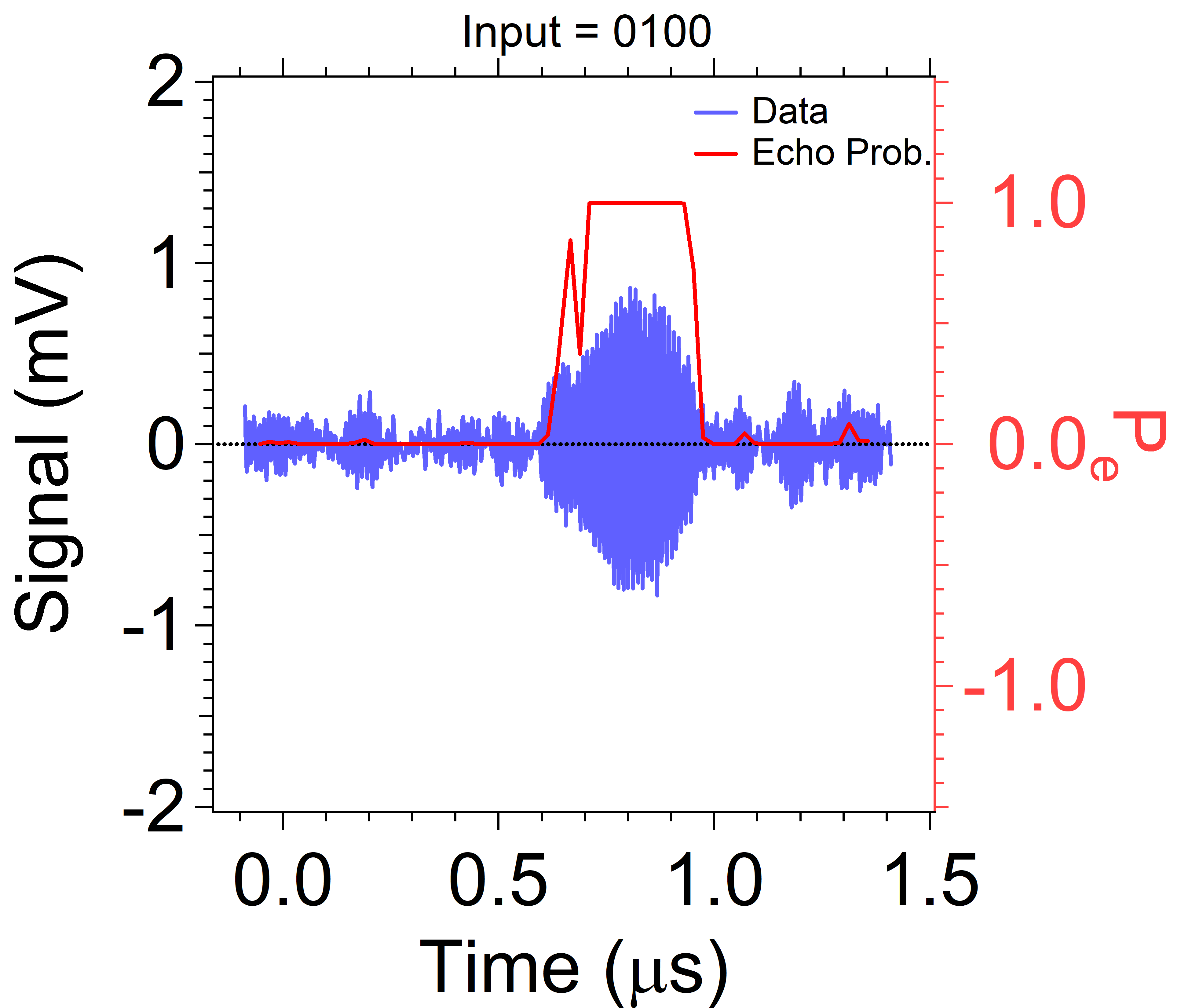}
\includegraphics[width=0.24\textwidth]{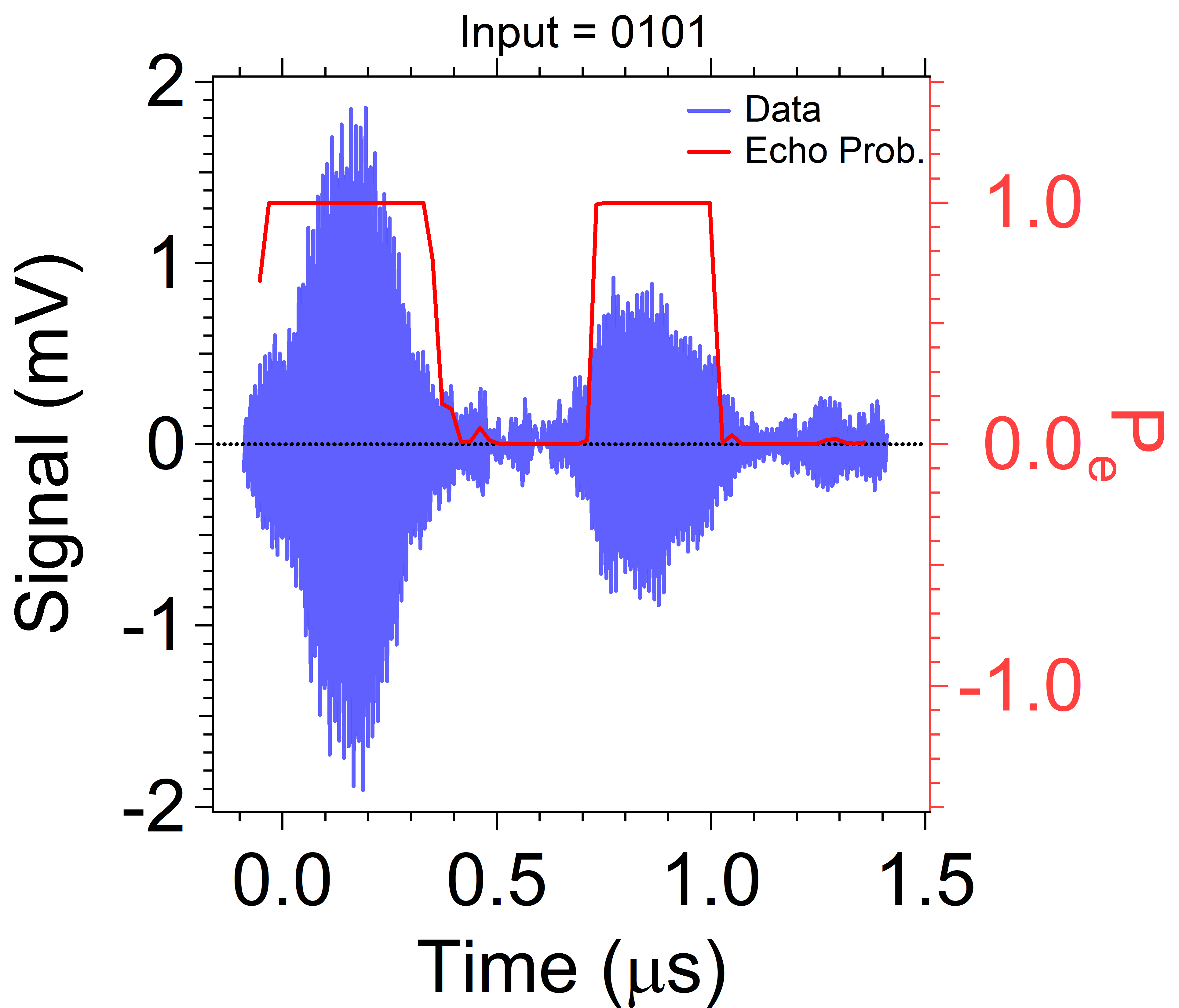}
\includegraphics[width=0.24\textwidth]{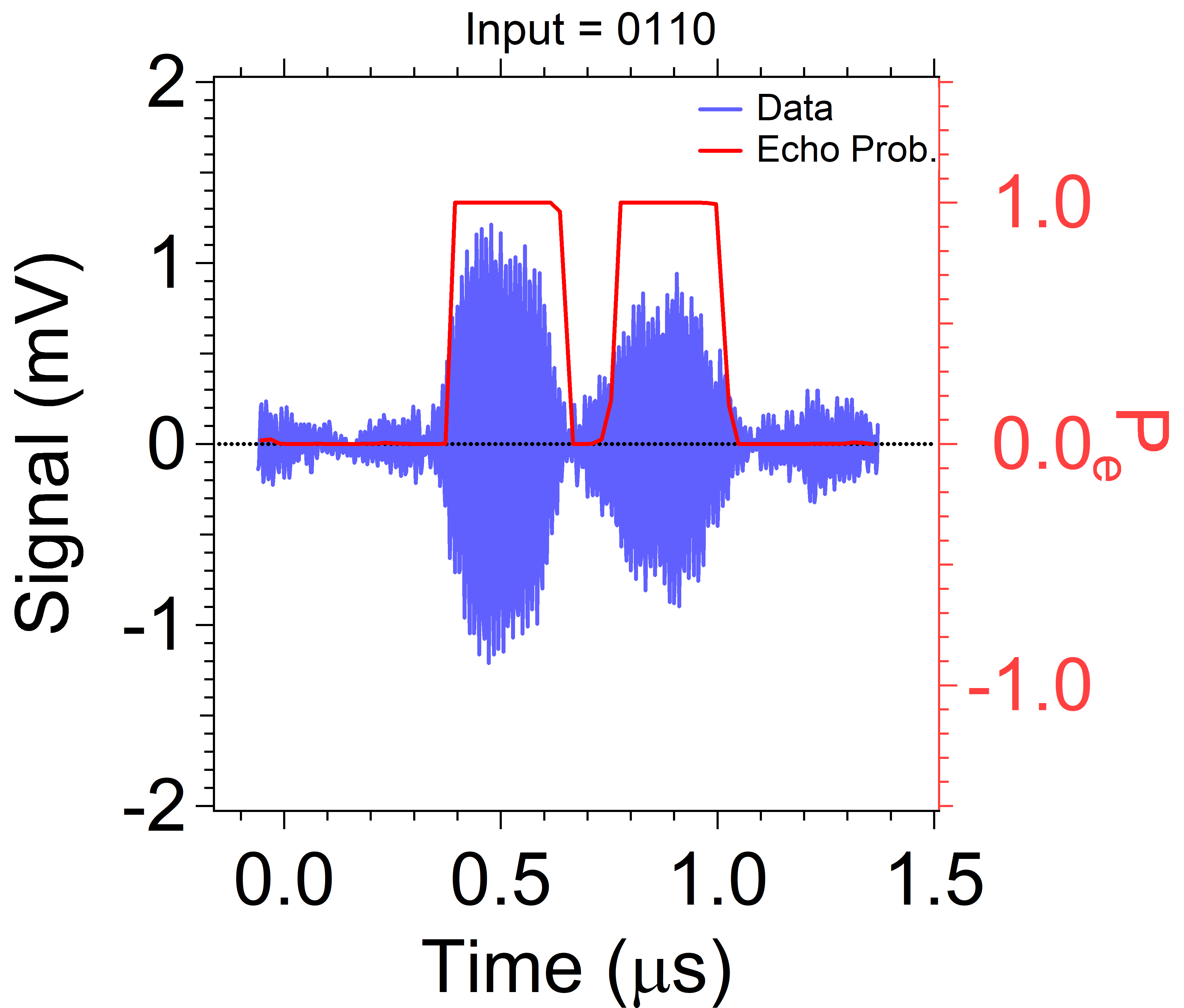}
\includegraphics[width=0.24\textwidth]{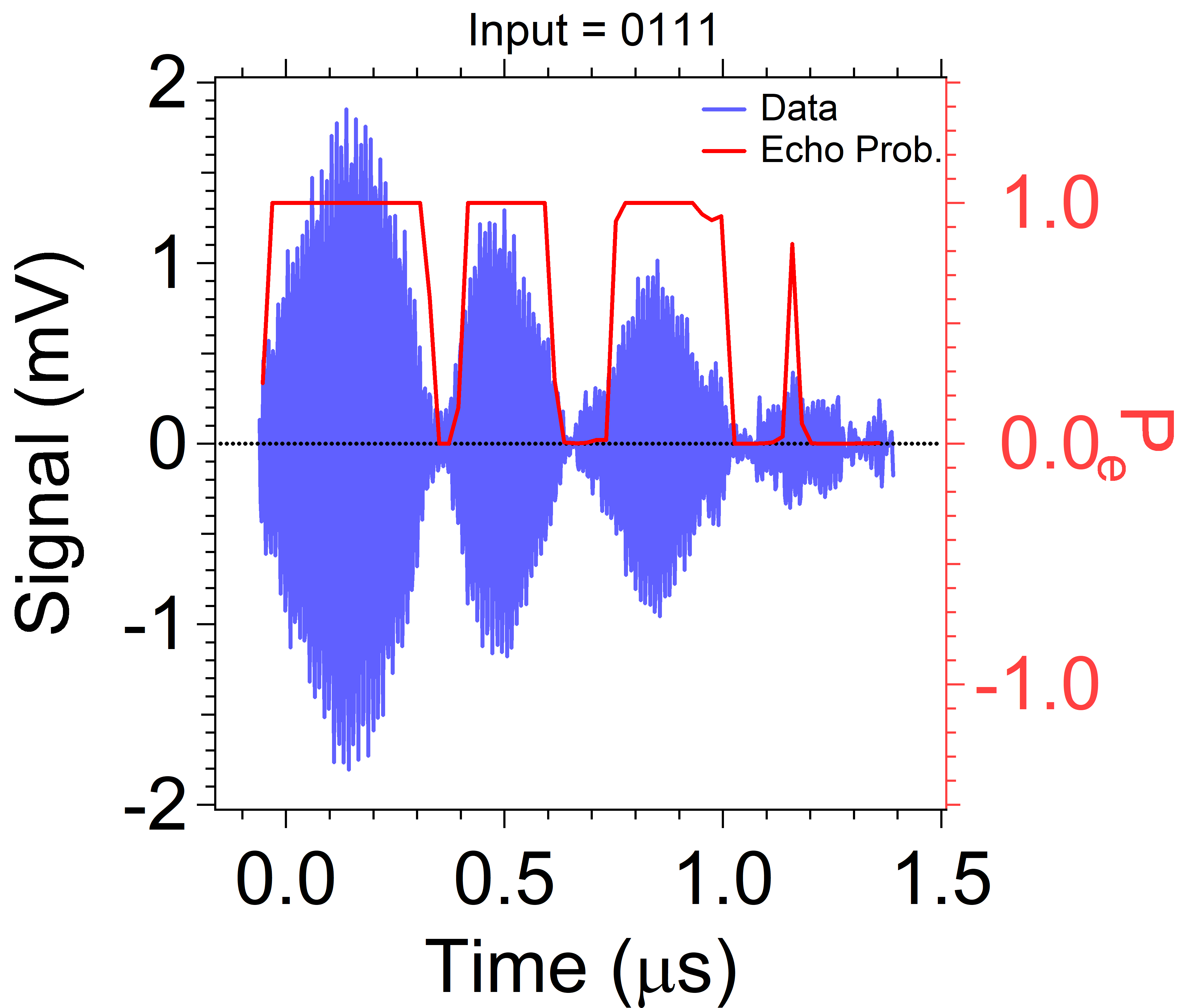}\\
\includegraphics[width=0.24\textwidth]{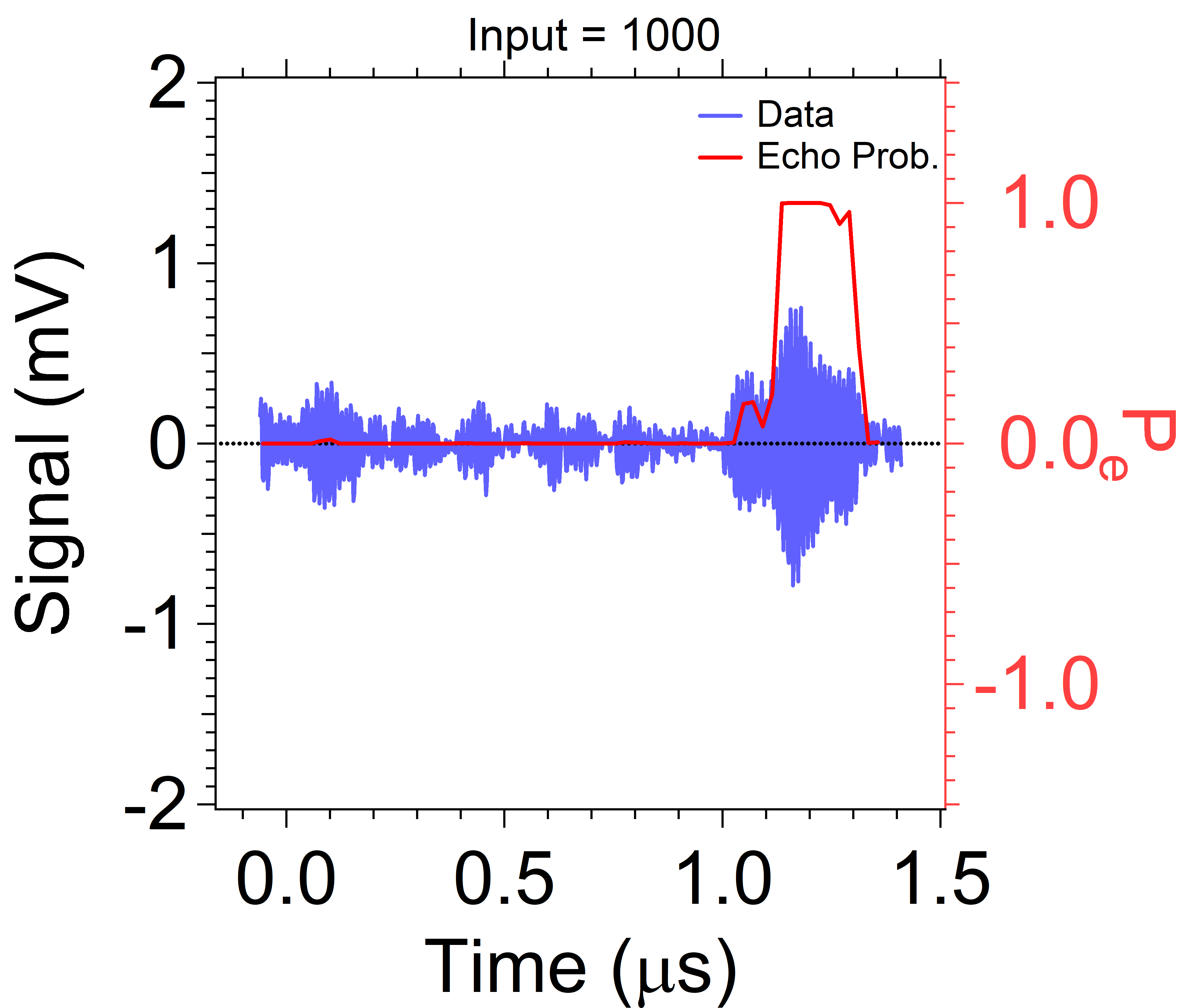}
\includegraphics[width=0.24\textwidth]{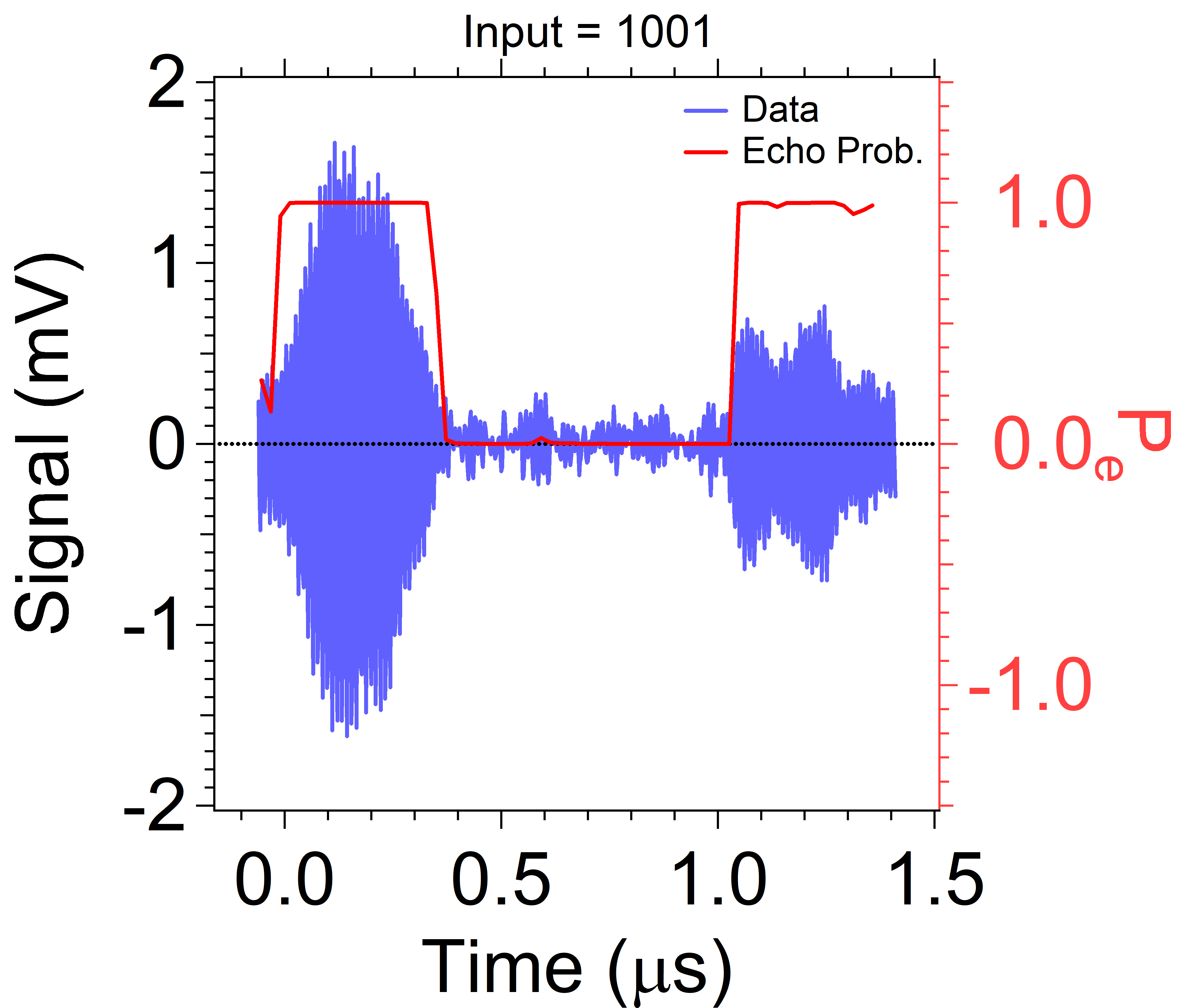}
\includegraphics[width=0.24\textwidth]{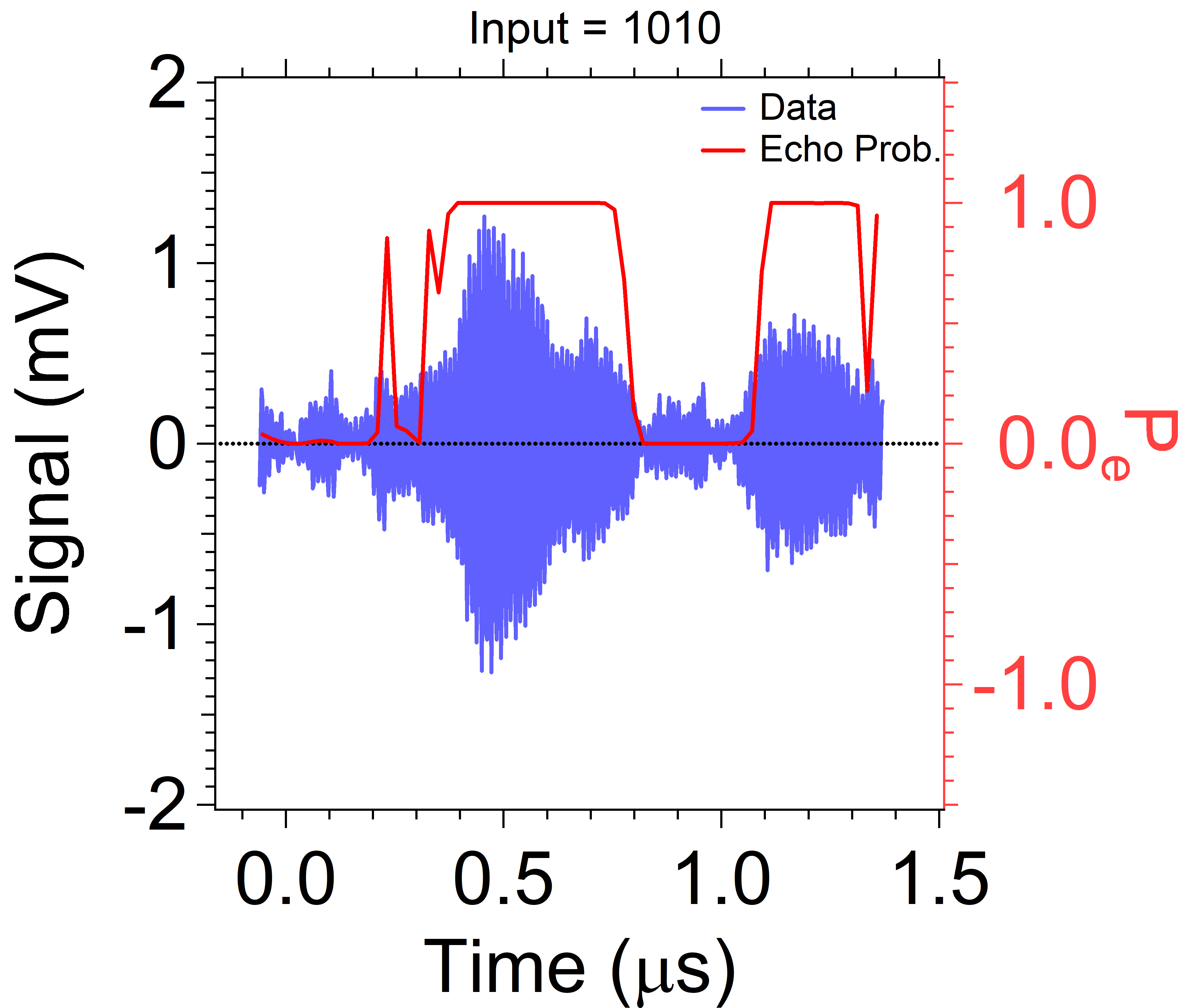}
\includegraphics[width=0.24\textwidth]{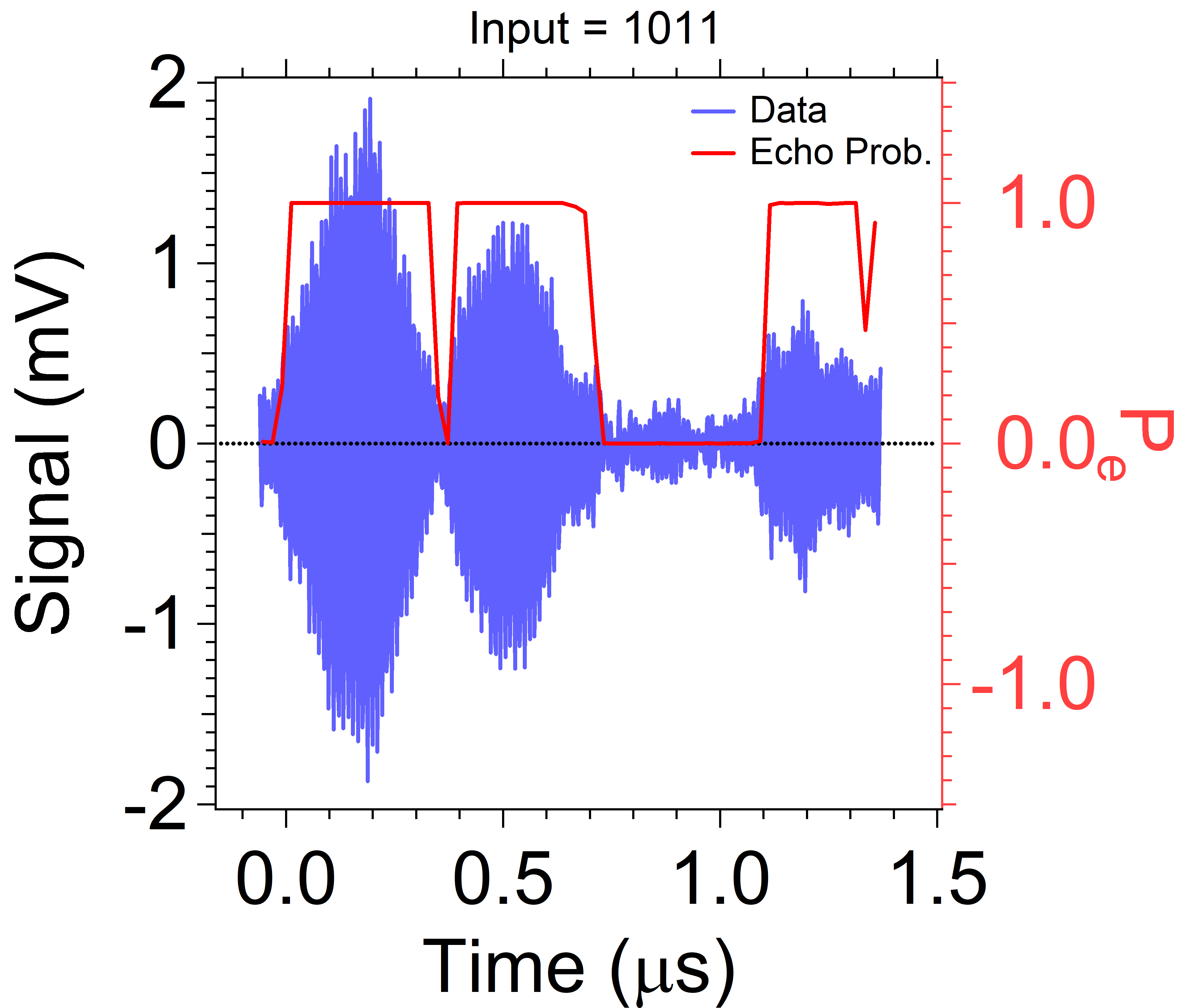}\\
\includegraphics[width=0.24\textwidth]{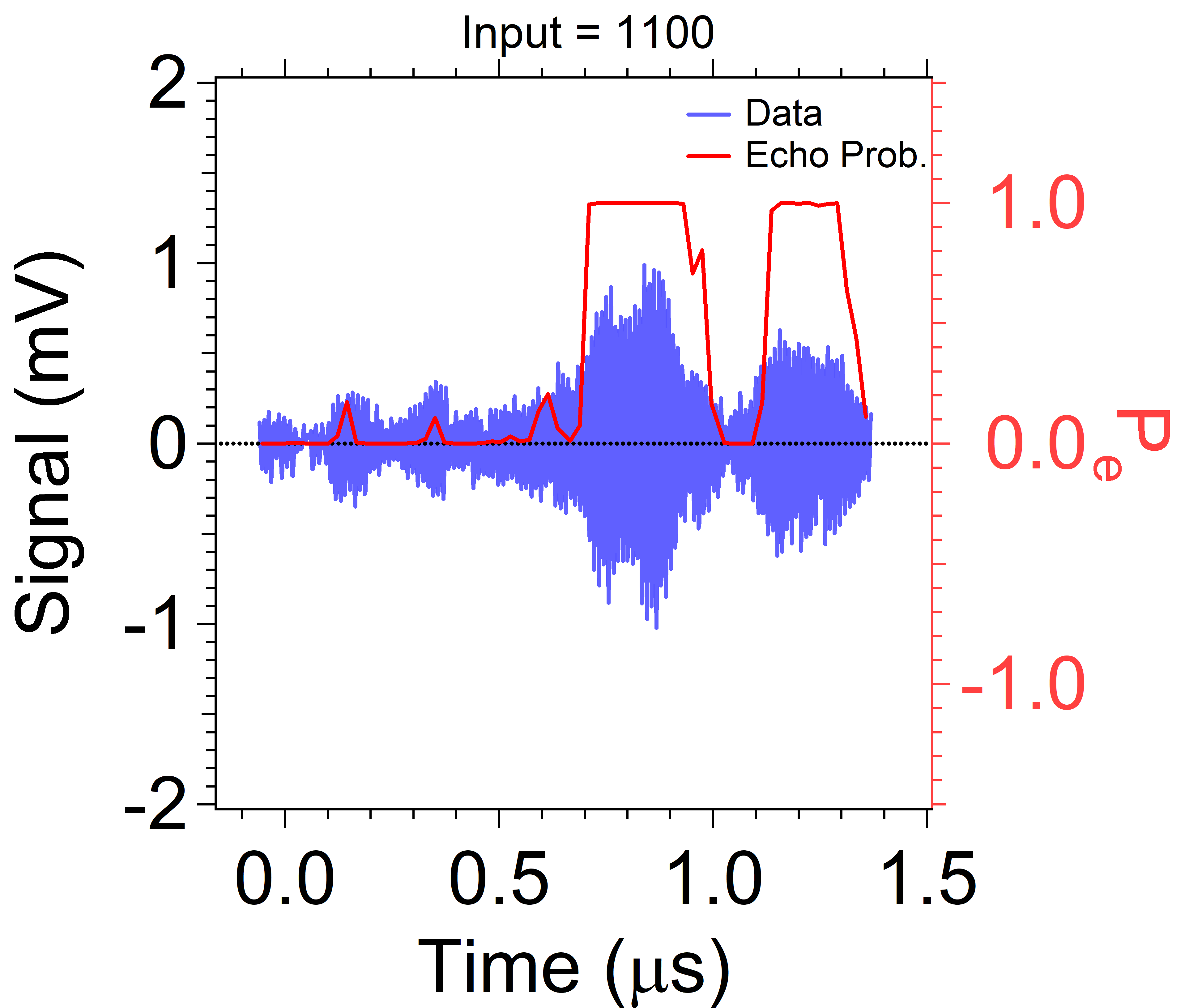}
\includegraphics[width=0.24\textwidth]{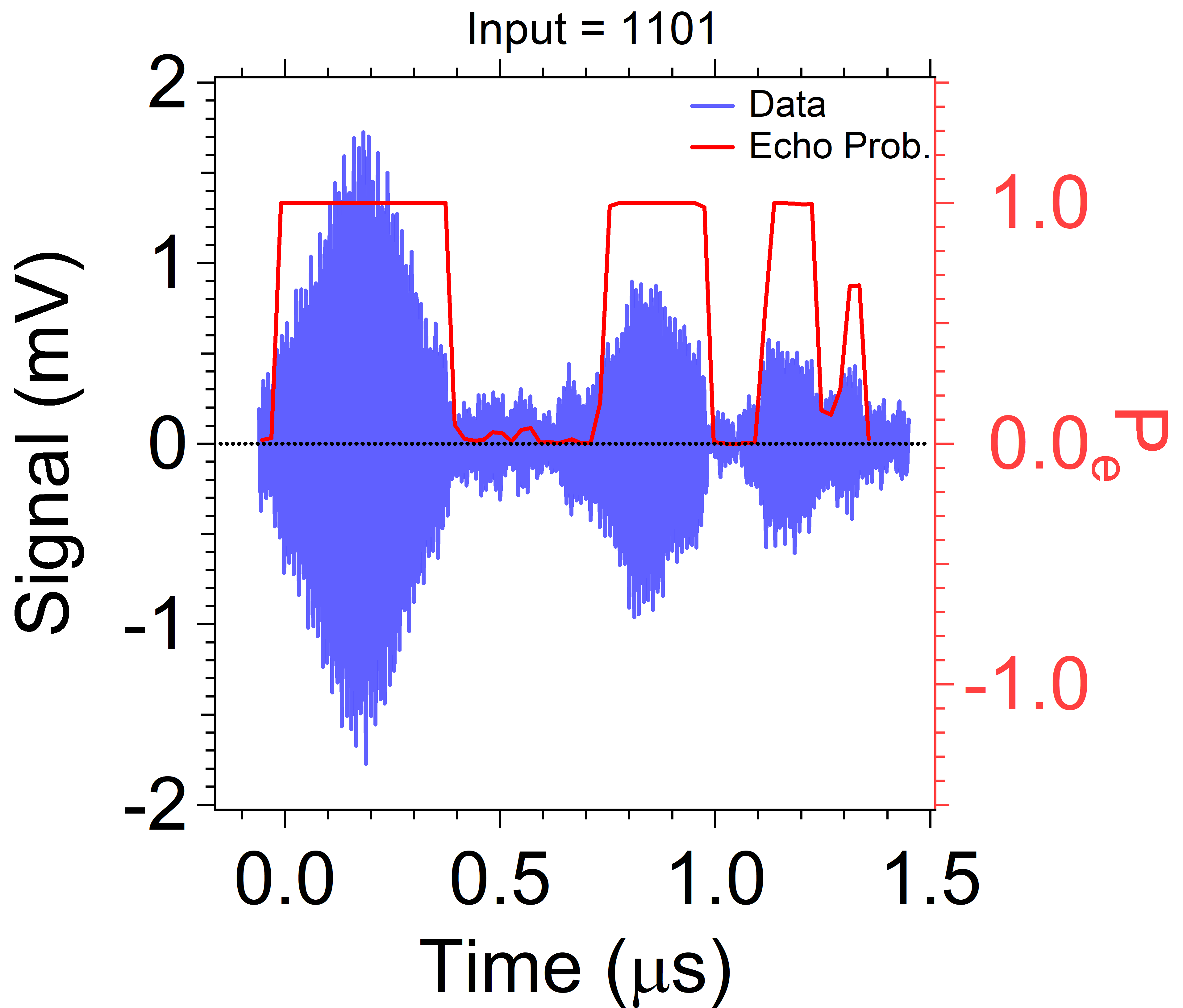}
\includegraphics[width=0.24\textwidth]{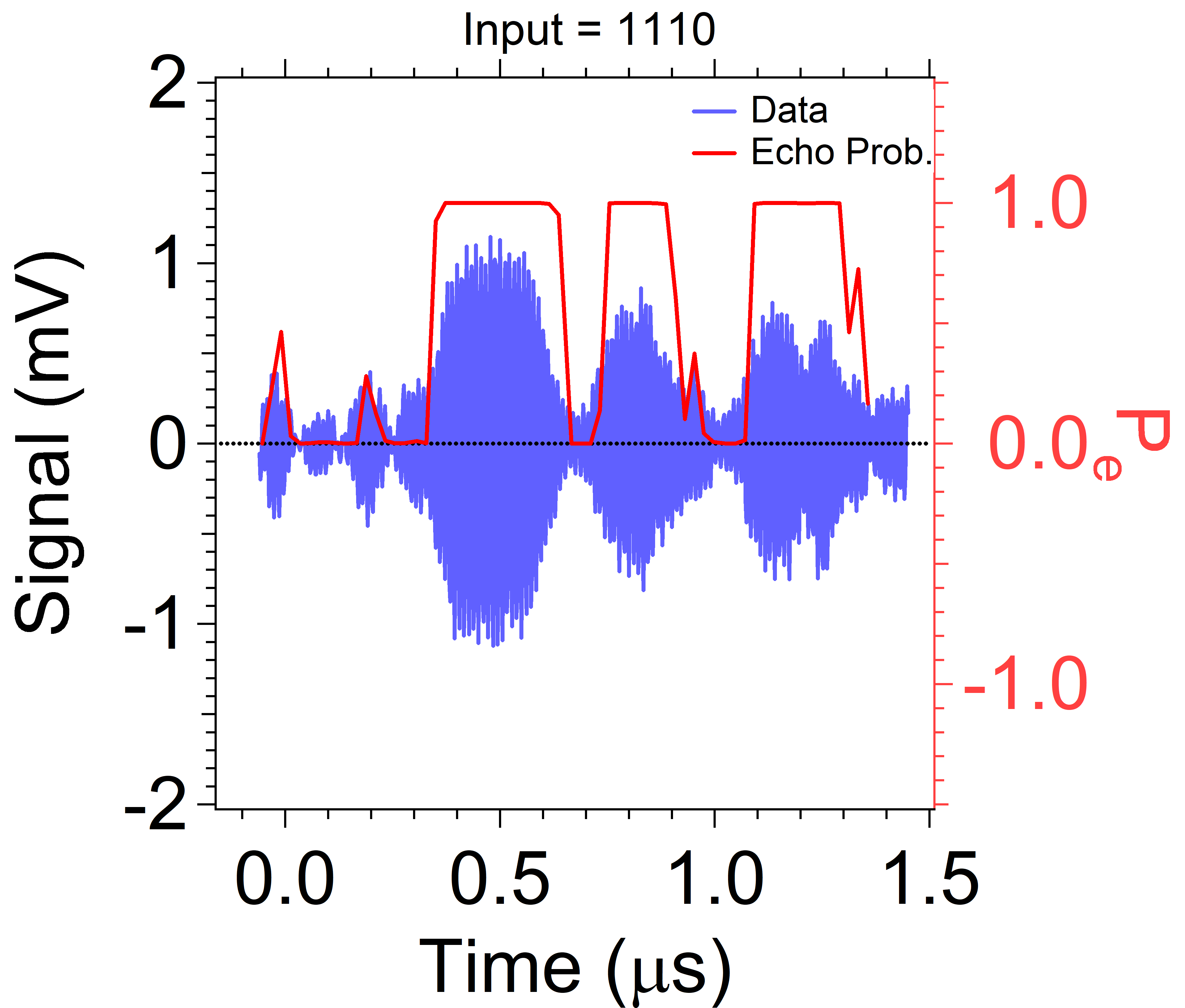}
\includegraphics[width=0.24\textwidth]{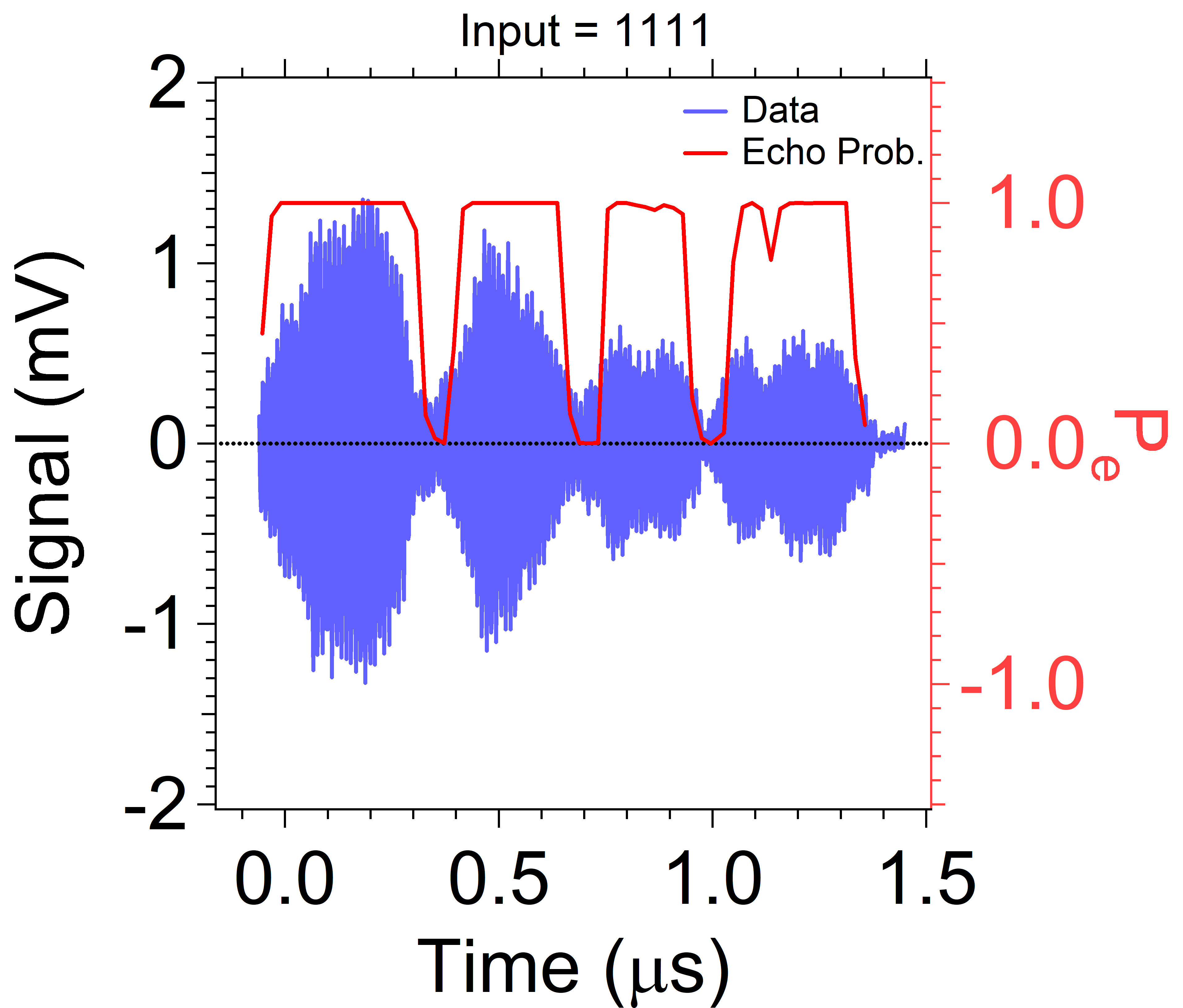}
\caption{Machine Learning-assisted recognition of output echoes of a Storage/Retrieval protocol. Blue traces are the measured raw outputs from all the possible sequences of 4 classical bit inputs sequences (being 1 = pulse ON and 0 = pulse OFF, from "0000" or $j=0$ at top left down to "1111" or $j=15$ at bottom right, see labels), while red traces are the probability, $p_{e}$, that the ANN gives in classifying the output as an echo. The probability scale $p_{e}$ is added on the right axis in red color for better comparison and clarity.}
\label{fig_stor_retr_echo}
\end{figure}

We first consider the Storage/Retrieval protocol as implemented in \cite{bonizzoniNPJQUANT2020}. We generate and acquire all the $2^4=16$ sequences arising from 4 input pulses, which allow us to codify and store into the ensemble all the decimals numbers between 0 and 15 in binary units, as shown in Fig. \ref{fig_stor_retr_echo}. Hereafter we will refer to the bit positions as $i$ (with $i=1,\dots,4$ in input order), and to the decimal numbers codified in input as $j$ (with $j=0,\dots,15$, which will also correspond to the number of the sequence itself). 
Each sequence gives a different train of output echoes, in which their order is reversed with respect to the input one according to the time reversal given by the $\pi$ pulse \cite{bonizzoniNPJQUANT2020}. Moreover, the amplitude of the output echo train decays over time according to the phase memory time of the ensemble \cite{bonizzoniNPJQUANT2020}. 
Each raw trace is sliced and sequentially given as input to the Neural Network (see Supplementary Information for details) in order to predict the corresponding echo probability, $p_{e}$, for each portion of the trace. The result for each raw trace is shown in Fig. \ref{fig_stor_retr_echo} (red traces and axes) together with its corresponding raw measured signal (blue traces). Here, the probability scale (in units) is shown on the right axis for better comparison. We notice that a plateau of probability with value $p_{e}\approx\,1$ is clearly visible in correspondence of each output echo and that the value is zero when there is no echo. All echo signals are correctly identified for all sequences. This is remarkable if one considers that no prior information on the positions or on the number of echo expected was given as input or during the training of the network.         

\onecolumngrid\
\twocolumngrid\

\begin{figure}[h!]
\centering
\includegraphics[width=0.38\textwidth]{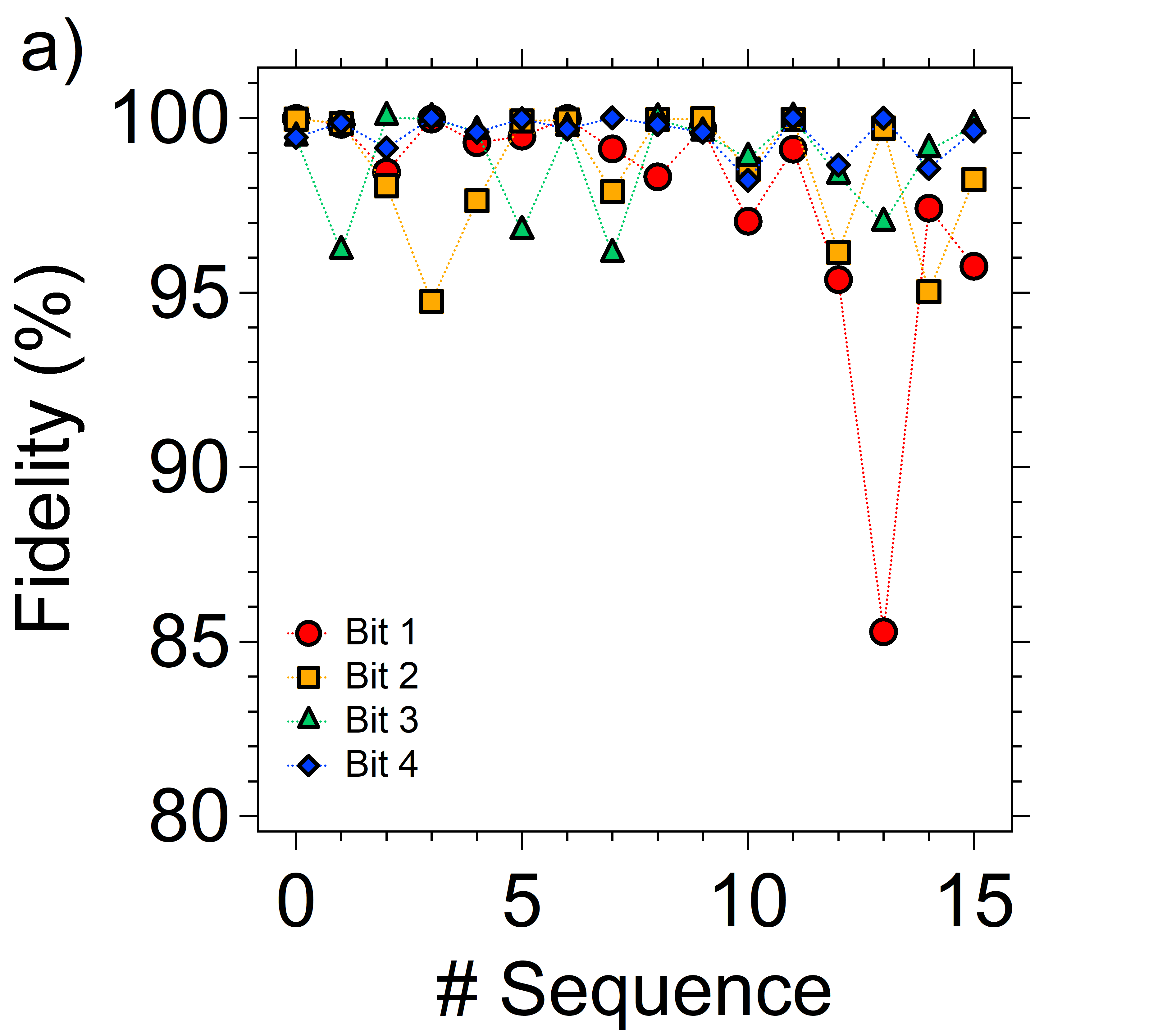}
\includegraphics[width=0.38\textwidth]{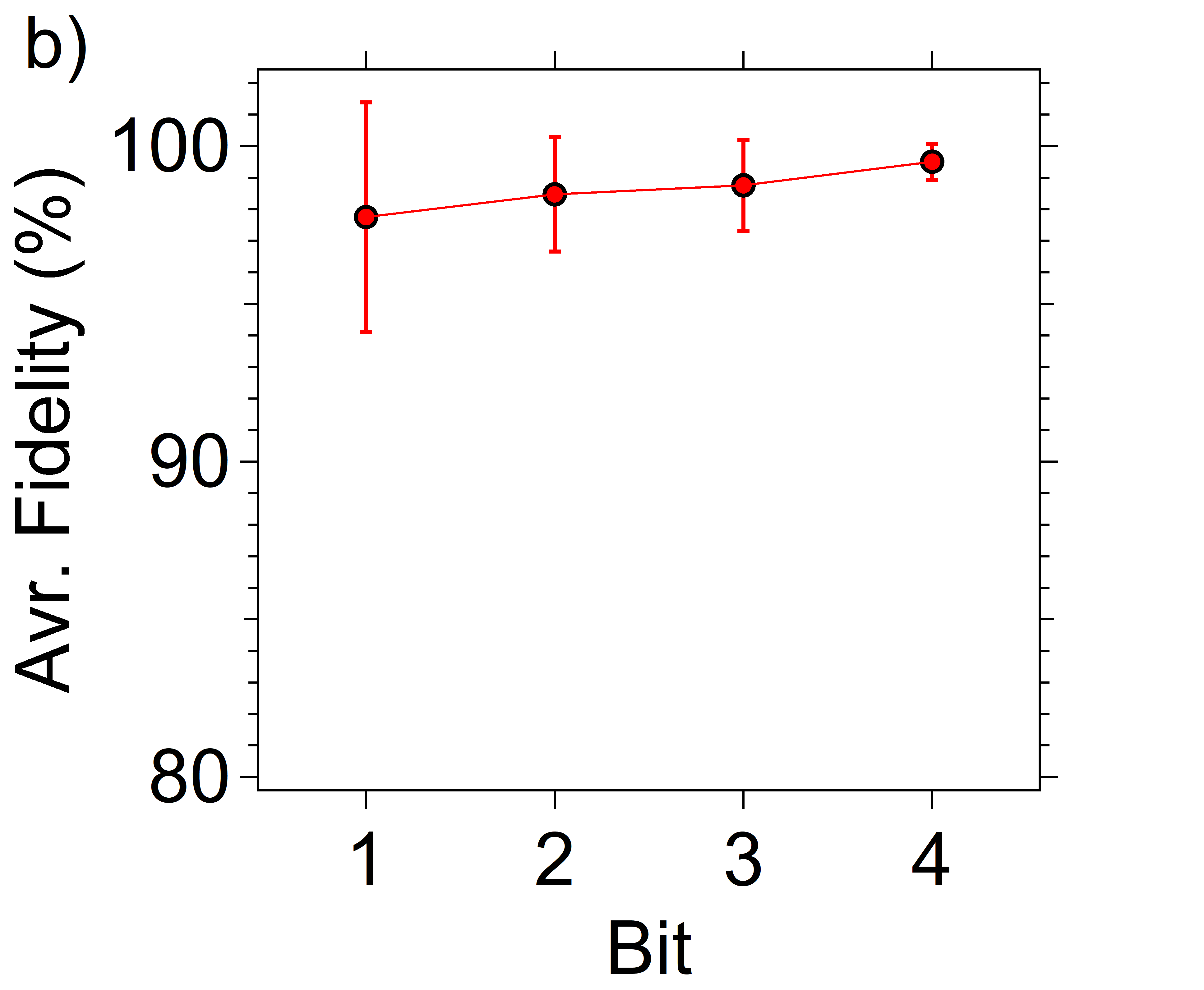}
\caption{Fidelity of bit recognition after clusterization of the echo probability (red traces in Fig. \ref{fig_stor_retr_echo})  according to the definition of Eq. \ref{eq_fidelity}. a) Fidelity of each bit (from 1 up to 4 in input order) as a function of the input bit sequence. b) Average Fidelity per bit resulting from the average of all Fidelity values per bit in a). Error bars are calculated as standard deviation of each corresponding point.}
\label{fig_stor_retr_fid}
\end{figure}

We further analyze these results by inferring the sequences given as input. To this end, we first split the output trace given by the ANN in four, equally-spaced windows. Then, we apply K-means clustering method on each of them to analyze the aggregation of the echo probability in two clusters. The probability value assigned for each window is the one of the centroid of the cluster with the larger number of points (See Supplementary Information for details). This method allows us to infer the bit value for each of the four logical positions of each sequence. 
To quantify the accuracy of the recognition we define Fidelity as in Eq. \ref{eq_fidelity}. This corresponds to one minus the absolute value of the difference between the expected nominal value of the bit with position $i$ 
for a given $j^{th}$ sequence 
, $A_{i}^{j}$, and the corresponding echo probability obtained after clusterization, $p_{e,i,rev}^{j}$. These latter values are reversed (subscript "\emph{rev}") for a given $j$ to take into account the inversion due to the $\pi$ pulse. We express the Fidelity in percentage units and, according to this definition, the larger the value the better is the accuracy of the inference (\textit{i.e.}, 0\% is a failure in recognition while 100\% is perfect agreement). The Fidelity obtained with Eq. \ref{eq_fidelity} are shown in Fig. \ref{fig_stor_retr_fid}. 
\begin{equation}
F_{i}^{j} = (1 - \left| A_{i}^{j}  - p_{e,i,rev}^{j} \right|)\cdot 100 \%
\label{eq_fidelity}
\end{equation}
The value is always above 95\% for each bit and trace (the unique exception is the point $i=1,j=13$, for which the value is $F_{i=1}^{j=13}\approx\,85\,\%$), suggesting that each bit value is correctly assigned in all cases and with high confidence. Best Fidelity values are obtained for $i=4$ because this is the last bit stored into the ensemble and the first to be retrieved. This means it is the one having largest amplitude and, hence, the easiest to be detected and inferred. Conversely, the worst Fidelity is obtained for $i=1$, that is the last echo to be retrieved and, hence, the one with the smaller output amplitude. Comparable trends are obtained for $i=2$ and $i=3$. In Fig. \ref{fig_stor_retr_fid} we show the average Fidelity for each $i^{th}$ bit over all its corresponding $j$, $\hat{F}_{i}=(\sum_{j=0}^{15}F_{i}^{j})/16$. The average Fidelity is always $\hat{F}_{i}\geq97\,\%$, reaching up to $\hat{F}_{i=4}\approx\,99\,\%$. We attribute the increase of the average Fidelity to the effect of the phase memory time and to the different total precession time at which each bit is retrieved, as discussed above.

\begin{table*}
\begin{ruledtabular}
\begin{tabular}{p{2cm}|p{2cm}|p{2cm}|p{2cm}|p{2cm}|p{2cm}|p{2cm}}
\multicolumn{2}{c|}{Method} & \multirow{2}{*}{ \makecell{\% success}} & \multirow{2}{*}{$\hat{F}_{1}$} & \multirow{2}{*}{$\hat{F}_{2}$} & \multirow{2}{*}{$\hat{F}_{3}$} & \multirow{2}{*}{$\hat{F}_{4}$} \\
 Echo/Noise & Bit Inference &   &  &  &  &  \\
\hline
ANN$^{ML}$ & K-means$^{ML}$ & 100 & 97.7$\pm$ 3.6 & 98.4$\pm$ 1.8 & 98.7$\pm$ 1.4 & 99.5$\pm$ 0.6 \\
    & Average & 90.6 & 84.7$\pm$ 19.1 & 83.3$\pm$ 14.0 & 89.4$\pm$ 8.9 & 96.5$\pm$ 4.0 \\
    & Max Search & 84.4 & 94.6$\pm$ 18.8 & 75.9$\pm$ 42.8 & 72.8$\pm$ 43.8 & 92.7$\pm$ 16.4 \\
    & Find Peaks & 90.6 & 93.8 $\pm$ 21.4 & 87.4$\pm$ 34.0 & 95.6$\pm$ 15.3 & 89.0$\pm$ 29.5 \\
\hline
\multicolumn{2}{c|}{Find Peaks} & 98.4 & 93.7 $\pm$ 25.0 & 100 & 100 & 100 \\
\end{tabular}
\end{ruledtabular}
\caption{Summary of the percentage of success and of the average Fidelity per bit ($\hat{F}_{i}$) for different methods used in this work (see text and Supplementary Information). The percentage of success is defined as the ratio between the number of cases for which $F_{j}^{i}\geq\,70$\% and the total number of bit values (64). Superscript "ML" denotes Machine Learning methods.}
\label{tab_comparison_stor_retr}
\end{table*}

To better asses our results, we compare our method with conventional scripts, which are not based on ML (see Supplementary Information for details). In particular, we first apply a standard algorithm for peak search to the raw traces of Fig. \ref{fig_stor_retr_echo}, in which a voltage threshold is defined by the user to attribute a measured signal to noise or echo. Our results suggest that the choice of the threshold can affect the Fidelity of the recognition especially for the smallest signals ($i=1$) and that, under our conditions, a perfect inference (\textit{i.e.} correct identification of all bits for all sequences) cannot be achieved (with the best threshold value giving 1 error over 64 cases, see Supplementary Information). The comparison with Fidelity obtained with ML suggests that these latter method can have two advantages with respect to conventional (\textit{i.e.} non-ML) ones: i) no threshold values need to be manually added and, ii) more accurate results are found for signals which are small and having amplitudes more comparable to noise.\\ 
We then repeat a similar comparison on the post-selection method used on the probability $p_{e}$ (red traces of Fig. \ref{fig_stor_retr_echo}) to infer the bit values. We evaluate the Fidelity for several different conventional search algorithms (see Supplementary Information). Here in particular none of the non-ML method gave perfect inference. The comparison further suggests that Clusterization can give more accurate results and lower variability in the predictions and that it is also less sensitive to errors correlated to signal variations or to the definition of the windows. The percentage of success and the average Fidelity per bit obtained with the methods considered in this work are summarized in Tab. \ref{tab_comparison_stor_retr}.

\subsection{Machine Learning-Assisted Phase Detection}
\label{sec_ml_phase}

In this section we focus on the recognition of the phase of the output echo measured after an Hahn echo sequence. The same resonator, sample and experimental set up of Sec. \ref{sec_ml_amplitude} are used. Here, the phase of the first $\pi/2$ pulse, $\phi_{\pi/2}$, or of the second $\pi$ pulse, $\phi_{\pi}$, can be tuned between 0 and $360^{\circ}$ while monitoring the output echo quadrature signals. The phase control on the $\pi/2$ pulse allow us to initialize the precession of the spins with any initial arbitrary angle in the xy precession plane, while the control on the phase of the $\pi$ pulse allows us to change the direction of the spin refocusing and, hence, to add an arbitrary phase shift during precession.   
We develop an ANN taking as input the output echo traces (which are the output signals from channels I and Q of the mixer, respectively) and giving the corresponding phase value as output. The data used for training the network were taken by measuring the echo during a full phase sweep of the first $\pi/2$ pulse between 0 and $360^{\circ}$ in steps of 2 degrees. The data sets used to test the predictions of the network are different from the training ones, but they were measured under similar experimental conditions.

\begin{figure}[h!]
\centering
\includegraphics[width=0.35\textwidth]{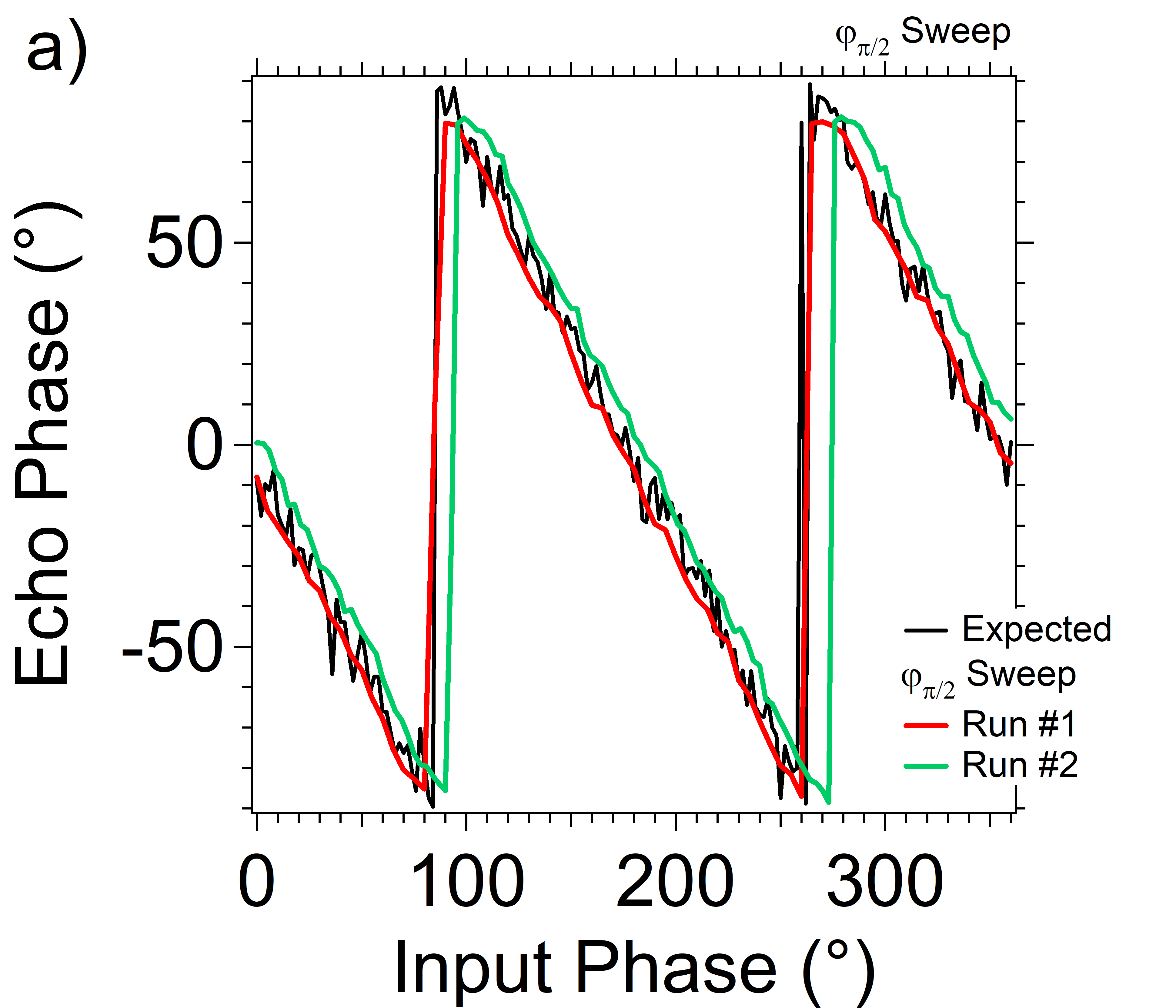}
\includegraphics[width=0.35\textwidth]{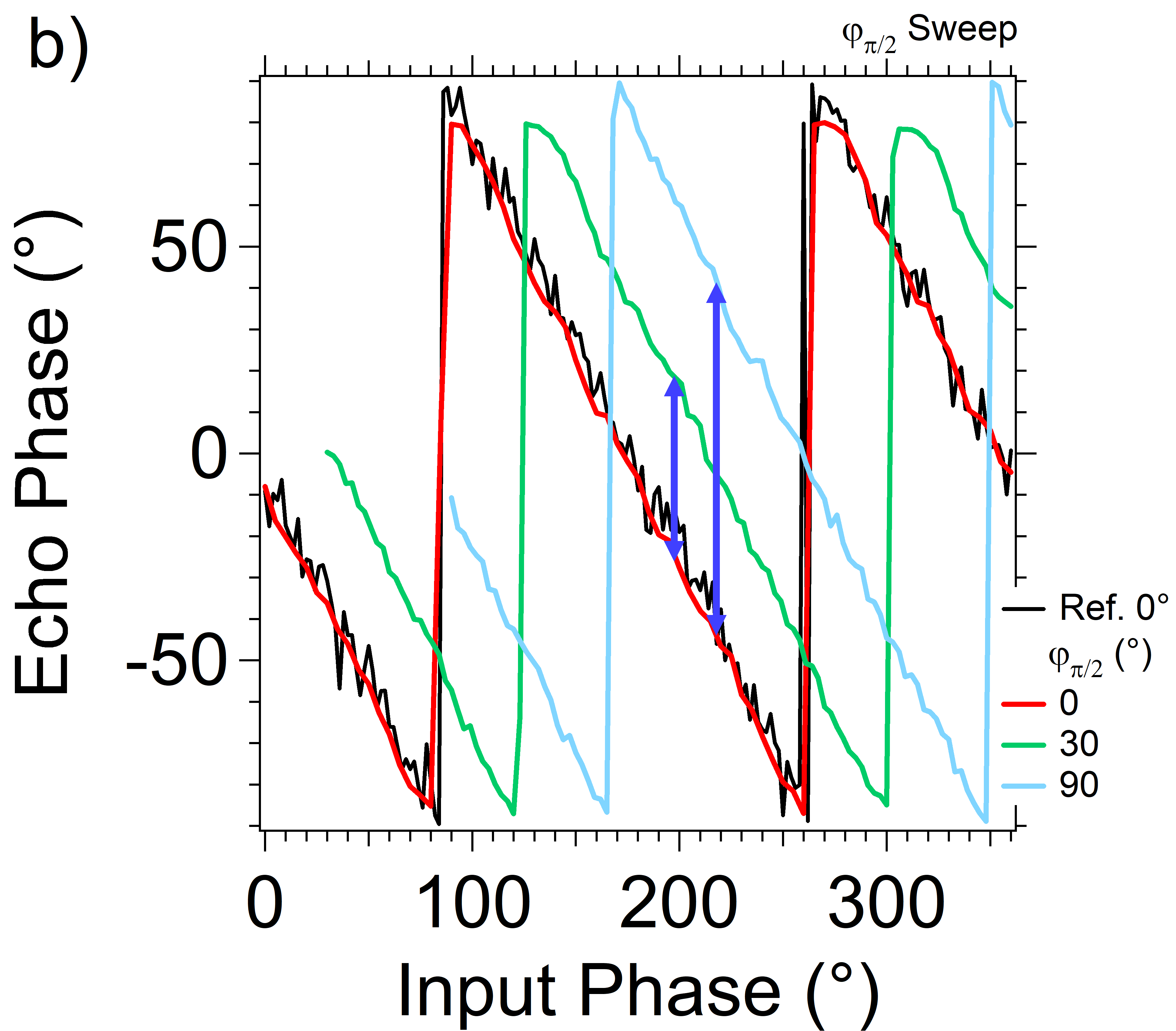}
\caption{Machine Learning-assisted recognition of the phase of the output Hahn echo. a) Inferred phase as a function of the input phase given to the $\pi/2$ pulse for two different data sets (Run \#1 and \#2). Black trace shows the values expected from the analysis of the output quadratures. b) Inferred phase as a function of the input phase sweep of the $\pi/2$ pulse when an initial phase bias (up to $90^{\circ}$, see legend) is added. Vertical blue arrows help in identifying the additional relative phase shift.}
\label{fig_phase_rec_firstpulse}
\end{figure}

We first check the output phase predicted by the network with two different data sets in which the phase $\phi_{\pi/2}$ pulse is swept between 0 and $360^{\circ}$, but with two different step sizes (3$^{\circ}$ and 5$^{\circ}$, Run \#1 and \#2 in Fig. \ref{fig_phase_rec_firstpulse}, respectively). The phase predicted by the network agrees with the one expected from the analysis of the echo quadrature signals (black trace in Fig. \ref{fig_phase_rec_firstpulse}, see Supplementary Information). Moreover, also the $180^{\circ}$ periodicity of the phase is correctly recognized. These results demonstrate the capability of ANNs in detecting the phase of a spin echo. Here we note that there is a small difference of $\approx7-10^{\circ}$ between the two different runs (\#1 and \#2). This small bias is consistent with the uncertainty given by the variability (noise) of the expected theoretical behaviour (black trace). To check the effect of the different spin initialization on the predictions, we repeat the recognition with new data sets in which the initial value of the $\phi_{\pi/2}$ sweep is different from zero. Results are shown in Fig. \ref{fig_phase_rec_firstpulse}.b. The values expected without considering the initial shift are added for comparison. The network correctly recognizes the periodical oscillations of the phase and also an additional phase bias introduced in the echo signal. Since the initial phase of the magnetization is expected to be preserved on the characteristic time scale of the memory time during precession, we can attribute the phase bias found to the effect of the different initialization performed by the $\pi/2$ pulse. This suggests that the network is sensitive to the different initial condition of the magnetization and, more generally, to an additional phase shift introduced in the precession by means of a single microwave pulse.\\

\begin{figure}[h!]
\centering
\includegraphics[width=0.4\textwidth]{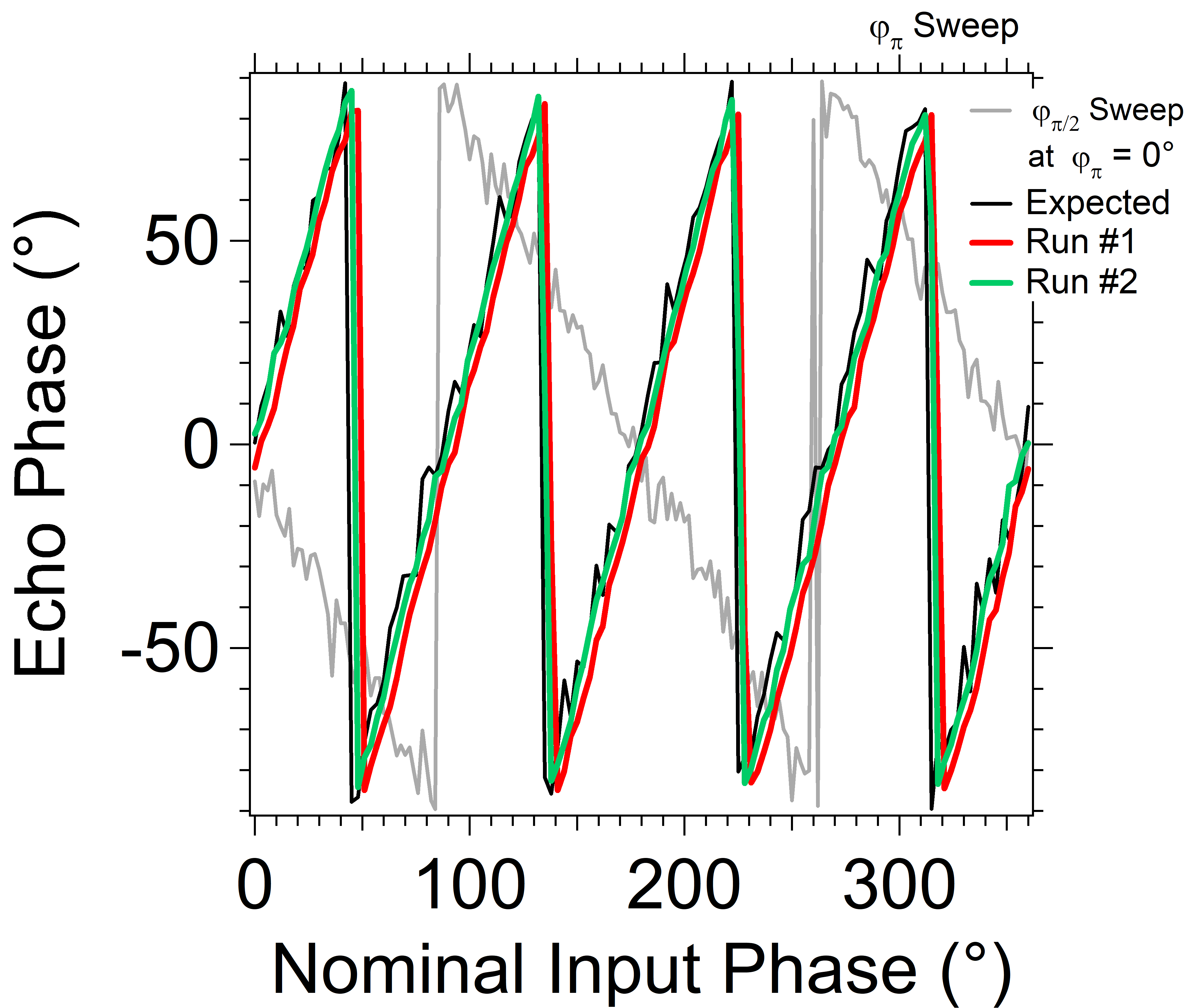}
\caption{Machine Learning-assisted recognition of the phase of the output Hahn echo when the phase given to the $\pi$ pulse is swept between 0 and 360$^{\circ}$ for $\phi_{\pi/2}=0$. Two different dataset taken under similar conditions (Run \#1 and \#2) are shown. Black trace shows the values expected from the analysis of the output quadratures, while gray trace shows the result expected from a phase sweep of the $\pi/2$ pulse at fixed $\phi_{\pi}=0$ for comparison.}
\label{fig_phase_rec_secpulse}
\end{figure}

We further inspect this latter effect using an additional data set in which the phase $\phi_{\pi/2}$ is fixed to zero and the phase of the $\pi$ pulse, $\phi_{\pi}$, is swept between 0 and 360$^{\circ}$. As mentioned above, this phase control changes the refocusing axis of the spins and introduces an additional phase shift in the precession, according to the relation $\phi_{fin}=2\phi_{\pi}-\phi_{in}$ (in which $\phi_{in}$ and $\phi_{fin}$ are the phases before and after the application of the pulse, respectively). Here, we remark that the training data set used with the network is the same as above, in which the phase of the Hahn echo is mapped as a function of $\phi_{\pi/2}$ for $\phi_{\pi}=0$. Results are shown in Fig. \ref{fig_phase_rec_secpulse}. The values expected from the analysis of the quadratures (black trace) and from a sweep of $\phi_{\pi/2}$ for $\phi_{\pi}=0$ (gray trace) are added for comparison. It is clear that the sweeps in $\phi_{\pi}$ have opposite sign with respect to the sweep of $\phi_{\pi/2}$ and that the period of the oscillations is halved (\textit{i.e.} doubled frequency). This behavior is consistent with the inversion of the precession given by the $\pi$ pulse and with the phase shift added, as expected from the term $-\phi_{in}$ and the prefactor 2 in $\phi_{fin}=2\phi_{\pi}-\phi_{in}$. These results shows that machine learning allows one to recognize the additional phase-control introduced by a single MW pulse during precession. This holds as far as phase control is performed using a single pulse and provided that the total number of pulses and the experimental conditions are not changed with respect to training ones.\\ 

\section{Discussion and Conclusions}
\label{sec_concl}

We have shown that it is possible to use ANNs to assist the manipulation and readout of molecular spin qubits embedded into planar microwave resonators. We first consider the output of a Storage/Retrieval protocol for sequences of 4 binary digits (\textit{i.e.} 16 possible decimal numbers), showing that the Network can recognize the position of the output echoes from the measured raw experimental traces. Here, we remark that no prior information on the number of echoes or on their position within the sequence was given to the ANN. This results hold potential for the analysis of traces containing an unknown number of echo signals, or which are heavily affected by noise or fluctuations. We then show that an additional clustering procedure (an unsupervised ML method) allows us to correctly infer the input sequence. To quantify the accuracy of the recognition we define a Fidelity (Eq. \ref{eq_fidelity}) which results to be above 95 \% (with lower bound of $\approx\,97\,\%$ for the average Fidelity). We then compare our method to different conventional algorithms which are not based on ML methods (see Supplementary Information). Our results suggest that ANNs can outperform the accuracy of non-ML methods when the measured signals are comparable to the noise level (\textit{i.e.} poor signal-to-noise ratio) and that they also require less operation by the user (\textit{e.g.} definition of thresholds). Furthermore, also the bit inference assisted by the Clustering can give more accurate results with respect to standard scripts and it is less affected by intrinsic signal fluctuations. 
We then consider the phase of the molecular spin qubit by focusing on the Hahn echo sequence, the simplest protocol allowing for the manipulation of the magnetization of the spin ensemble and for its readout. We first show that an ANN can successfully recognize the phase of the output echo from the analysis of raw experimental measured data (output I and Q channels of mixer). The network can also recognize additional single-pulse phase shifts introduced during the initialization of the precession ($\pi/2$ pulse) or during the refocusing ($\pi$ pulse) of the magnetization. Here we additionally notice that these latter results (Fig. \ref{fig_phase_rec_firstpulse} and \ref{fig_phase_rec_secpulse}) contain some special cases. In fact, the choice $\phi_{\pi/2}=0,180^{\circ}$ for the $\pi/2$ pulse gives a rotation by $\pi/2$ radiants about the $\pm \hat{x}$ axis, while $\phi_{\pi/2}=90,270^{\circ}$ gives an effect equivalent to a rotation by the same angle about the $\pm \hat{y}$ axis. These operations have the same effect of the application of the Rotational Gates $R_{x}(\theta),R_{y}(\theta)$ for the case $\theta=\pi/2$ \cite{krantzAPL2019,crooksTECHNOTE2022,marinescuACAMPRESS2012}. The choice $\phi_{\pi}=0,180^{\circ}$ for the $\pi$ pulse corresponds to a rotation having the same effect of a Pauli $\pm X$ gate, while the choice $\phi_{\pi}=90,270^{\circ}$ has the effect of a Pauli $\pm Y$ gate \cite{krantzAPL2019,crooksTECHNOTE2022,marinescuACAMPRESS2012,haroche}. These considerations suggest that our approach holds potential towards the implementation of ML-assisted single-gate operations such as Pauli Gates and Rotational Gates and, with the proper training and optimized network, it might be further extended to other gate operations.
Thanks to the relatively-small dimensions of the ANNs used, it runs on standard and commercially-available hardware and it requires limited computational resources (see Supplementary Information), which allows for an easy and light implementation also on small processors or instrumentation.
 
It is worth mentioning that our method could be further extended in the view of implementing molecular spin qudits \cite{morenopinedaCHEMSOCREV2018,gomezleonPRA2022,jenkinsPRB2017,hussainJACS2018,chizziniPHYSCHEMCHEMPHYS2022,gimenoCHEMSCI2021,chiccoCHEMSCI2021}. This constitutes a twofold problem: i) upgrade and optimization of our experimental set-up to allow for the realization of multiple tones spectroscopy (microwave or microwave plus radiofrequency) on molecular qudits, and ii) generalization of our ML learning methods to account for the larger complexity of the problem. While the first problem is technological and can be addressed with commercially-available instrumentation and solutions, the latter one requires a careful choice of the input and of the output parameters to be used and, consequently, of the corresponding training dataset and process. For instance, once a qudit protocol is defined and experimentally implemented, a possible approach might be adding an additional label on the training dataset to account for the frequency used.   

Although developed on molecular spin qubits, our results and method are based on the analysis of raw echo traces and does not require additional information on the sequence (\textit{e.g.} pulse durations, interpulse delay), on the physical system used (\textit{e.g.} memory time) or on experimental and measurement conditions (\textit{e.g.} integration time, number of avererages). Therefore, it can be extended to more complex PW sequences, such as Dynamical Decoupling Protocols \cite{bonizzoniNPJQUANT2020,wisePRXQUANT2021} or ENDOR-like scheme to perform Storage/Retrieval of information into nuclear spins \cite{wolfoewiczPRL2015,wuPRL2010}. More in general, it can be applied to other spin systems and to all systems showing quantum coherence, such as diluted magnetic centers \cite{mortonJMR2018}, superconducting qubits \cite{Kjaergaard2020,krantzAPL2019} and solid-state spin qubits based on semiconductor quantum dots \cite{chatterjeeNATPHYSREV2021}.

Finally, further extension and outlooks are also possible. In particular, the recognition of the echo signal from noise can find application in experiments in which the echo signals are expected to be rather small or heavily affected by noise, \textit{e.g.} using resonators with active volumes equal or below nanoliter \cite{bonizzoniAPPLMAGNRES2022,abhyankarREVSCIINSTR2022,abhyankarSCIADV2020,ranjanAPL2020} or addressing molecules integrated on surfaces \cite{yamabayashiJACS2018}. Our results on phase recognition can be applied also to the instantaneous phase value and not only to the average one (see Supplementary Information). Moreover, the introduction of the proper corrections during the training of the ANN would allow to automatically take into account any imbalance between the I and Q channels of the mixers \cite{sabahPARTACCPROC1998} or any additional phase shift or bias of the set-up. Another possibility consists in the implementation of similar ANNs within on-line methods, paving the way for real time signal recognition, eventually at single-shot measurement level. Lastly, further possibility is the implementation of adaptive automated optimization protocols \cite{poderiniPHYSREVRES2022,melnikovPRL2020}. This might allow for the automated calibration and optimization of complex pulse sequences, such as the above-mentioned Dynamical Decoupling \cite{bonizzoniNPJQUANT2020,wisePRXQUANT2021}, in analogy with a black box-like operation in which the response of the set up is partially or not fully known \cite{poderiniPHYSREVRES2022,melnikovPRL2020}. Beneficial effects can be foreseen also down to the quantum regime of the electromagnetic excitation driving the spins, in which an efficient design of quantum states and of gate operations in presence of noise is highly desirable \cite{doolittlearxiv2022}.

\begin{acknowledgments}
We thank Prof. Simone Calderara (Università di Modena e Reggio Emilia) and Prof. Fabio Sciarrino (Università di Roma La Sapienza), Prof. Nicolò Spagnolo (Università di Roma La Sapienza), Dr. Emanuele Polino (Università di Roma La Sapienza) for useful preliminary discussions. We thank Prof. Roberta Sessoli (Università degli Studi di Firenze) for useful discussion and for manuscript proofreading.\\
This work was funded by the H2020-FETOPEN ”Supergalax” project (grant agreement n. 863313) supported by the European Community.\\
\end{acknowledgments}

\bibliography{biblio.bib}

\end{document}